\documentclass[usenatbib]{mnras}

\usepackage{multirow}
\usepackage{longtable}
\usepackage[T1]{fontenc}
 \usepackage{ae,aecompl}
\usepackage{amsmath}
\usepackage{breqn}
\usepackage{graphicx}
\usepackage{textcomp}
\usepackage{subfigure}
\usepackage{amssymb}    
\PassOptionsToPackage{authoryear}{natbib}
\usepackage{natbib}
\bibpunct{(}{)}{;}{a}{}{,}
\PassOptionsToPackage{english}{babel}
\usepackage{babel}

\renewcommand{\deg}{^\circ}

\def\deg {$^\circ$}
\def\flux {\mbox{erg~cm$^{-2}$~s$^{-1}$}}
\def\lum {\mbox{erg~s$^{-1}$}}

\newcommand{\xmm}{{\em XMM-Newton}}
\newcommand{\swift}{{\em Swift}}
\newcommand{\fermi}{{\em Fermi}}

\PassOptionsToPackage{hyphens}{url} 
\begin{document}

\title[Reverberating and superefficient nebulae]{Towards observing reverberating and superefficient pulsar wind nebulae}
\author[Torres, Lin, \& Coti Zelati]{Diego F. Torres$^{1,2,3}$, Tingting Lin$^{1,2}$, \& Francesco Coti Zelati$^{1,2}$ \\
$^1$Institute of Space Sciences (ICE, CSIC), Campus UAB, Carrer de Magrans s/n, 08193 Barcelona, Spain\\
$^2$Institut d'Estudis Espacials de Catalunya (IEEC), 08034 Barcelona, Spain\\
$^3$Instituci\'o Catalana de Recerca i Estudis Avan\c{c}ats (ICREA) Barcelona, Spain
}

\date{}
\pagerange{\pageref{firstpage}--\pageref{lastpage}} 
\pubyear{2018}

\maketitle

\label{firstpage}

\begin{abstract}
In a recent work, we numerically studied the radiative properties of the reverberation phase of pulsar wind nebulae 
(PWNe), i.e., when the reverse shock created by the supernova explosion travels back towards the pulsar, compressing the wind bubble. 
We focused on several well-characterized PWNe and used them as examples for 
introducing the concept of superefficiency.
The latter is a period of the PWN evolution, happening within reverberation, where the luminosity in a given band 
exceeds the spin-down power at the time. 
Here, we explore a broad range of PWN models 
to study their reverberation and superefficiency phases in a systematic way.
Armed with these models we consider two aspects: 
On the one hand, we analyze via Monte Carlo simulations how many Galactic PWNe are expected to be reverberating
or in a superefficiency stage at any given time,
providing the first such estimations. 
On the other hand, we focus on searching for observational signatures of such periods.  We analyze archival observations and check for the existence of
possible candidates for superefficient PWNe. We also
provide predictions for the future evolution of the magnetar nebula J1834.9-0846 (which we consider to be starting its reverberation period) 
along the next 50 years.
Using our simulations as input we study how sensitive current and future X-ray satellites (like {\it eXTP} or {\it Athena}) 
will be to observe such evolution, concluding that they will be able to track it in detail.

\end{abstract}


\section{Introduction}

Reverberation is a short but important phase in the evolution of all pulsar wind nebulae (PWNe), 
produced when the reverse shock created by the supernova explosion travels back towards the pulsar, compressing the wind bubble.
Its impact on the PWN evolution has been studied by
\cite{Gelfand2009,Vorster2013,Bandiera2014,Bucciantini2011,Martin2016,Torres2017,Torres2018b}, 
although it is not usual that radiative models used to interpret observations take it into account yet (see \cite{Gelfand2017} for a review).
When reverberation happens, the PWN size decreases, the magnetic field increases, and the medium transfers energy to the PWN
rather than the other way around.
Neither the instantaneous spin-down power of the pulsar nor the integrated rotational power up to this point in the nebula evolution
is the energy reservoir in this period.
The transfer of energy occurs via adiabatic heating of the particles in the nebula. Upgraded in energy, particles emit 
at higher frequencies. 
Because of the size reduction and magnetic field increase, the suddenly heated particles encounter appropriate 
conditions for radiative cooling via synchrotron and self-synchrotron emission.
This leads to a quick burn out of the particle population
together with a concurrent increase of luminosity.

In \cite{Torres2018b}, we showed that even the Crab nebula, associated to the more energetic pulsar of the sample then considered, 
will have its X-ray luminosity exceeding the spin-down power at a certain time of its future evolution.
We dubbed this period in the life of nebulae as the superefficient stage and characterized its main properties.
However, a systematic exploration of the superefficiency and reverberation periods was still lacking.
Also lacking was an estimation of the number of systems that could be in such superefficient, or in general, in a reverberation period, at any given time in our Galaxy.
These two tasks are undertaken in this work.
In order to achieve these results, we produce  
a set of systematic, and equally detailed PWNe simulations, and compare their reverberation periods 
as a function of the pulsar energetics and their different environmental properties. 
Using these PWN  evolutionary tracks, as well as current knowledge of the pulsar population 
and their birth rate, we compute the number of reverberating and superefficient PWNe via Monte Carlo  
simulations.
The final part of this work deals with the observational signatures of such periods. 
We analyze archival observations to see whether possible candidates of superefficient PWNe could have been already observed.
We also provide predictions for the future evolution of the  nebula around the magnetar
J1834.9-0846, which is considered to be starting its reverberation period (see \citet{Torres2017}),
along the next 50 years.
Using our simulations as input we finally study how well future X-ray satellites (like {\it eXTP} or {\it Athena}) 
will be able to observe such evolution.

\section{Systematic exploration of the superefficiency and reverberation periods}

When compared to one another, real PWNe  are subject to different environmental properties.
In order to compare generic reverberation and superefficiency properties for different pulsars in a more sensitive way,  
here we shall simulate putative PWNe generated by pulsars that differ from one another only in their energetics. 
Particularly,
we shall assume that the values of the initial spin-down power, $L_0$, of these putative pulsars are a multiple of that shown by a specific {\it real} pulsar anchor (e.g., the Crab pulsar),
but that they have the same braking index, $n$,  initial spin-down age, $\tau_0$, and environmental variables of the latter.  
%
%
Then, we shall obtain the spin-down power, $L_{sd}(t)$, and relate the  initial spin-down age with the  
 characteristic age, $\tau_c(t)$, as a function of time, as follows
  \begin{eqnarray}
  L_{sd}(t) &=& L_0 \left(1+\frac{t}{\tau_0} \right)^{-\frac{n+1}{n-1}} , \\
  \tau_0 &=&  \frac{2 \tau_c (t) }{ n-1} - t
    \end{eqnarray}
Additionally using the equations defining the characteristic age and spin-down  \citep{Pacini1973,Gaensler2006},
  \begin{eqnarray}
   \tau_c (t) &=& \frac{ P(t) }{ 2 \dot{P}(t) }, \\
    L_{sd}(t) &=& \frac {4\pi^2I {\dot{P}(t)}}{P(t)^3},
    \label{ePdot}
  \end{eqnarray}
%
we can derive the time evolution of $P$ and $\dot P$:
  \begin{eqnarray}
    P(t) &=& \sqrt{2 \pi^2\frac{I}{\tau_c(t) L_{sd}(t)}} ,\\
    \dot{P}(t)&=& \frac 12  \frac1{\tau_c(t)} \sqrt{\frac{2\pi^2 I}{\tau_c(t) L_{sd}(t)}}.
  \end{eqnarray}
These equations will track the evolution of the pulsars in the $P \dot P$-diagram (see Fig. \ref{PPdot} and Table 1).
PWNe simulations are done using the code Tide 2.3, presented in \cite{Martin2016}. 
Their radiative content can be found in \cite{Martin2012}, details on 
the treatment of the nebula magnetic field in \cite{Torres2013}, and of the initial free expansion phase as well as its systematic application
to young PWNe in \cite{Torres2014}. 
{  In particular, we are using a broken power law to describe the injected particles and consider that the
variation of the magnetic energy in the PWN is balanced by the fraction
of the rotational energy that powers the magnetic field (instantaneous power into the field is given by $\eta L(t)$, where $\eta$ is the magnetization)
and the adiabatic losses due to the expansion of the PWN. See equations 4 and 11 of  \cite{Martin2016}.}

\subsection{Details on the dynamical evolution}

Given the importance of the assumptions related to how the radius is computed along the dynamical evolution, we shall briefly summarize here the prescription introduced by
\cite{Martin2016} included in our PWN model, see also \cite{Gelfand2009} for a similar treatment.

Assuming that $v$, $R$,
$M$ and $P$ are the velocity, radius, mass and pressure in the PWN bubble; and that $v_{ej}$, $\rho_{ej}$ and $P_{ej}$ correspond to the
values of the velocity, density, and pressure of the SNR ejecta at the position of the PWN shell,
we compute the radius of the PWN during the free expansion phase and
compression by solving the equations given by \citet{Chevalier2005}, using a similar prescription as in \citet{Gelfand2009}:
\begin{eqnarray}
\frac{d R(t)}{dt}&=&v(t),\\
M(t)\frac{d v(t)}{dt}&=&4 \upi R^2(t) \left[P(t)-P_{ej}(R,t) \right],\\
\frac{d M(t)}{dt}&=&4 \upi R^2(t) \rho_{ej}(R,t) (v(t)-v_{ej}(R,t)); \\ 
                        & &           \hspace {0.1cm} {\rm for} \hspace {0.1cm} v_{ej}(R,t)<v(t) ,\\ 
 \frac{d M(t)}{dt}&=& 0; \hspace {0.1cm} {\rm for} \hspace {0.1cm} v_{ej}(R,t)\ge v(t) 
\label{eqmass}
\end{eqnarray}
To account for the radiative (as well as other) losses by the particles inside
the PWN in the computation of the PWN pressure $P(t)$, we solve the diffusion-loss equation
\begin{equation}
\label{diffloss}
\frac{\partial N(\gamma,t)}{\partial t}=Q(\gamma,t)-\frac{\partial}{\partial \gamma}\left[\dot{\gamma}(\gamma , t)N(\gamma,t) \right]-\frac{N(\gamma,t)}{\tau(\gamma,t)},
\end{equation}
{  The terms on the right hand side take into account the radiative energy losses, the losses or energy gains by compression or expansion of the PWN, and the escape of particles (we assume Bohm diffusion), for which $
\tau(\gamma,t)$ is here the timescale}.
We solve Eq. \ref{diffloss} 
in each time-step and compute the total
energy contained in pairs, $E_p$, by doing the integral,
\begin{equation}
\label{epar}
E_p(t)=\int_{\gamma_{min}}^{\gamma_{max}} \gamma m_e c^2 N(\gamma,t) \mathrm{d}\gamma.
\end{equation}
The pressure contributed by particles is
\begin{equation}
\label{ppar}
P_p(t)=\frac{3(\gamma_{pwn}-1)E_p(t)}{4 \upi R(t)^3}.
\end{equation}
$\gamma_{pwn}$ is the adiabatic coefficient of the PWN relativistic gas, which is fixed as 4/3. 
The magnetic pressure is 
\begin{equation}
\label{pmag}
P_B(t)=\frac{B^2(t)}{8 \upi}.
\end{equation}
And finally, the total pressure is  
\begin{equation}
P(t)=P_p(t)+P_B(t).
\end{equation}

In this prescription, 
the $v_{ej}$, $\rho_{ej}$ and $P_{ej}$ change if the PWN shell is surrounded  by unshocked ejecta ($R < R_{rs}$), or by shocked ejecta ($R_{rs} < R < R_{snr}$,
being $R_{snr}$ the radius of the SNR). The initial profiles for the unshocked medium are assumed, following \citet{Truelove1999,Blondin2001}, as
\begin{equation}
\label{vej}
v_{ej}(r,t)=\left \{
\begin{array}{ll}
r/t & {\rm for}\;\;\; r < R_{snr},\\
0 & {\rm for}\;\;\; r > R_{snr},
\end{array}  \right .
\end{equation}
\begin{equation}
\label{rhoej}
\rho_{ej}(r,t)=\left \{
\begin{array}{ll}
A/t^3 & {\rm for}\;\;\; r < v_t t,\\
A(v_t/r)^\omega t^{\omega-3} & {\rm for}\;\;\; v_t t < r < R_{snr},\\
\rho_{ism} & {\rm for}\;\;\; r > R_{snr},
\end{array}  \right .
\end{equation}
\begin{equation}
P_{ej}(r,t)=0,
\end{equation}
where
\begin{equation}
A=\frac{(5\omega-25)E_{sn}}{2 \upi \omega v_t^5},
\end{equation}
\begin{equation}
v_t=\sqrt{\frac{10(\omega-5)E_{sn}}{3(\omega-3)M_{ej}}}.
\end{equation}
The parameters $E_{sn}$ and $M_{ej}$ are the energy of the SN and the total ejected mass during the explosion, respectively. In our model,
for simplicity, we assume $\omega=9$ as in \citet{Chevalier1992,Blondin2001,Gelfand2009} for a type II SN. 
We have however tested that changing this steep decline of the density by using another power law index, e.g., $\omega=6$ the start of the reverberation is slightly affected, but the extent of the
 compression is maintained.

When the PWN shell is surrounded
by the shocked medium, then we use the appendix of \citet{Bandiera1984} to obtain the $v_{ej}$, $\rho_{ej}$ and $P_{ej}$ profiles.

For the shock trajectories, we use the semianalytic model by \citet{Truelove1999} for a non-radiative SNR.

The rebounce after the compression is treated with an ad hoc pasting into a Sedov expansion, i.e., we consider that the PWN bounces and starts the Sedov phase when its pressure reaches the pressure of the SNR's Sedov solution and that at
this time, the evolution of the radius of the PWN follows the relation given in \citet{Bucciantini2011}
\begin{equation}
\label{sedov}
R^4(t_{Sedov})P(t_{Sedov})=R^4(t)P(t),
\end{equation}
where $P(t)=\rho_{ism} v_{fs}^2/(\gamma_{snr}+1)$ is the pressure in the SNR forward shock. The term $v_{fs}^2$ is the velocity of the
forward shock given by the Truelove \& McKee equations.

\subsection{Sample}

We shall consider here four different specific cases of pulsar anchors for generating the models:  the magnetar
J1834.9-0846 (referred to as J1834), Crab, G0.9+0.1 (G09), and G54.1+0.3 (G54).
 We shall especially consider initial spin-down powers ranging from $L_0 \sim 10^{35}$ to $10^{40}$ erg s$^{-1}$ (see below for further discussion on this point).
Table 1 of \cite{Torres2018b} gives the parameters for the anchor pulsar and PWN properties, e.g., their braking index and the observational values for the current $P$ and $\dot P$,
as well as resulting features of their evolution (e.g., the duration of reverberation and of any superefficiency found in the energy ranges explored).
Table 1 here gives the $L_0$ values considered for each of the simulated models based on these real PWNe, whereas all other parameters of the simulated PWNe are
assumed as the corresponding original ones quoted in Table 1 of \cite{Torres2018b}). {  The latter also applies to fitted parameters of the real PWNe, like the magnetization
or the break energy.}

\begin{figure*}
   \centering
    \includegraphics[width=0.45\textwidth]{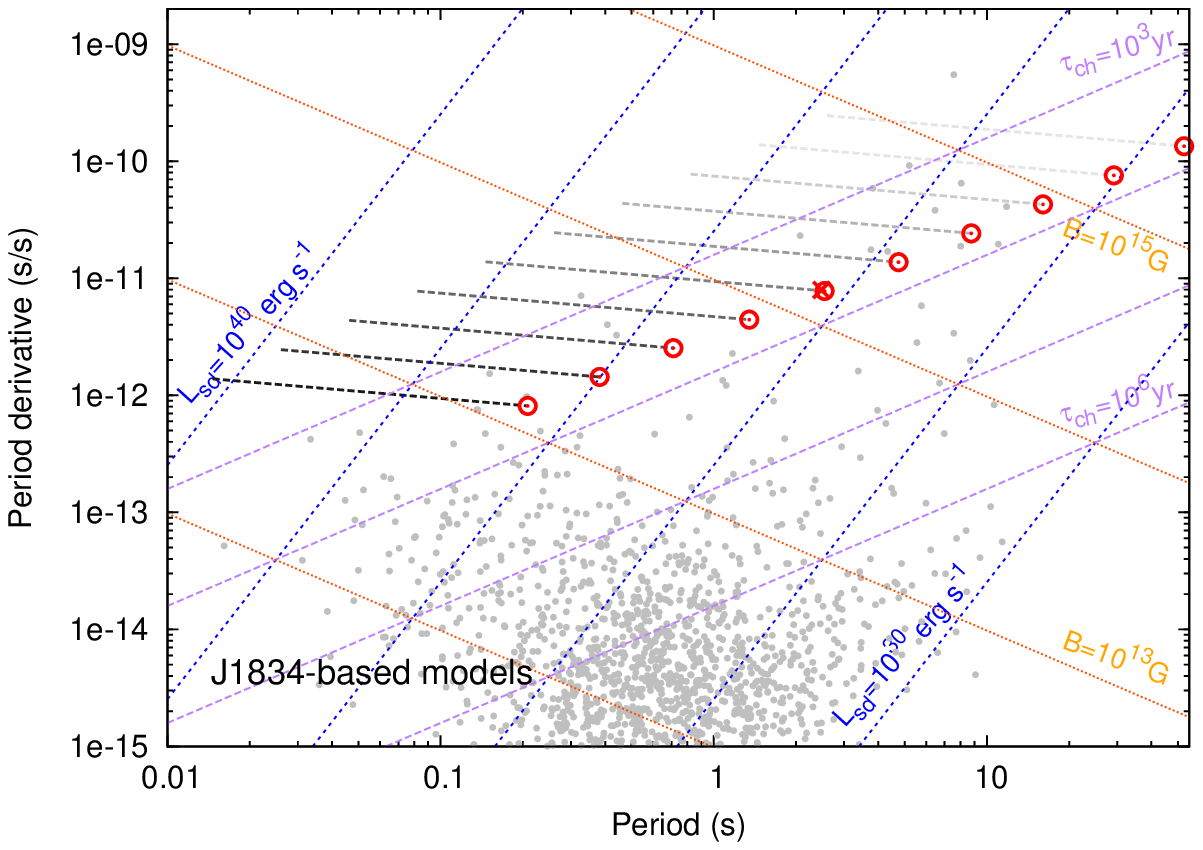}
        \includegraphics[width=0.45\textwidth]{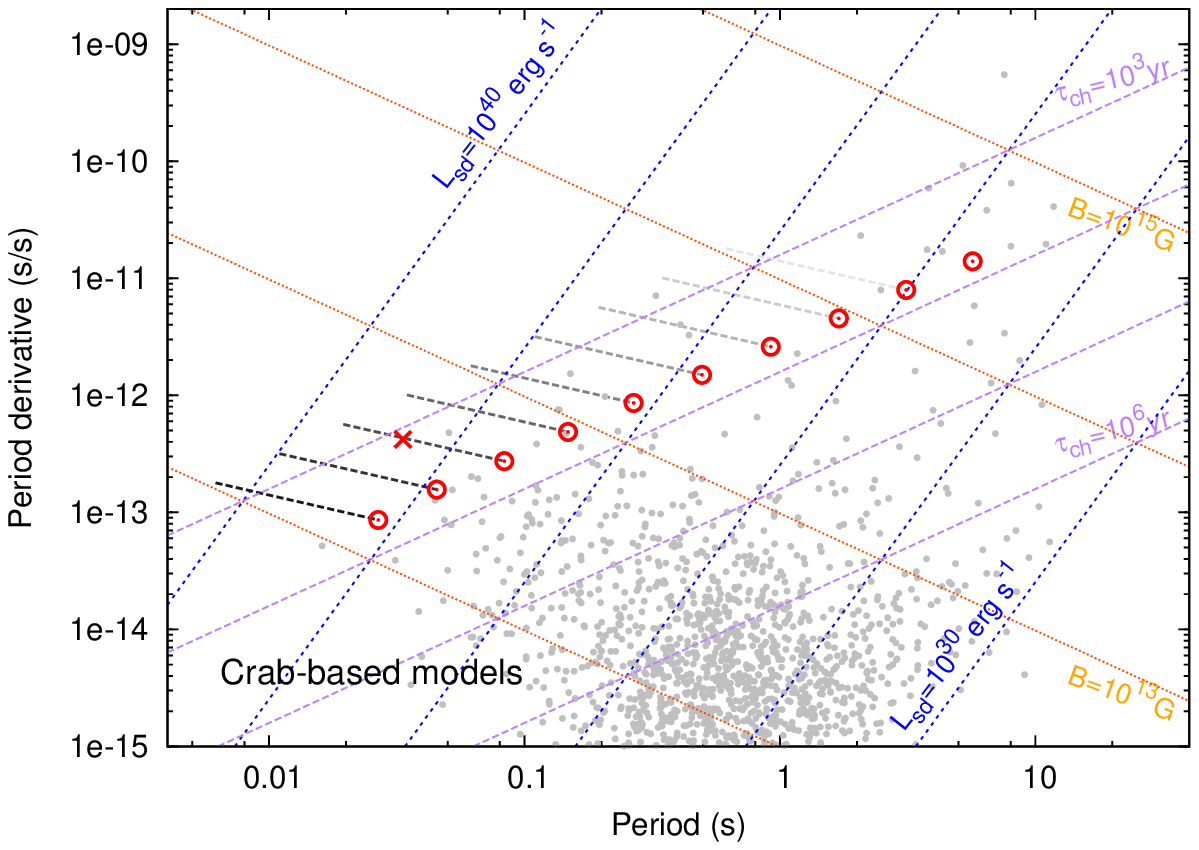}\\
            \includegraphics[width=0.45\textwidth]{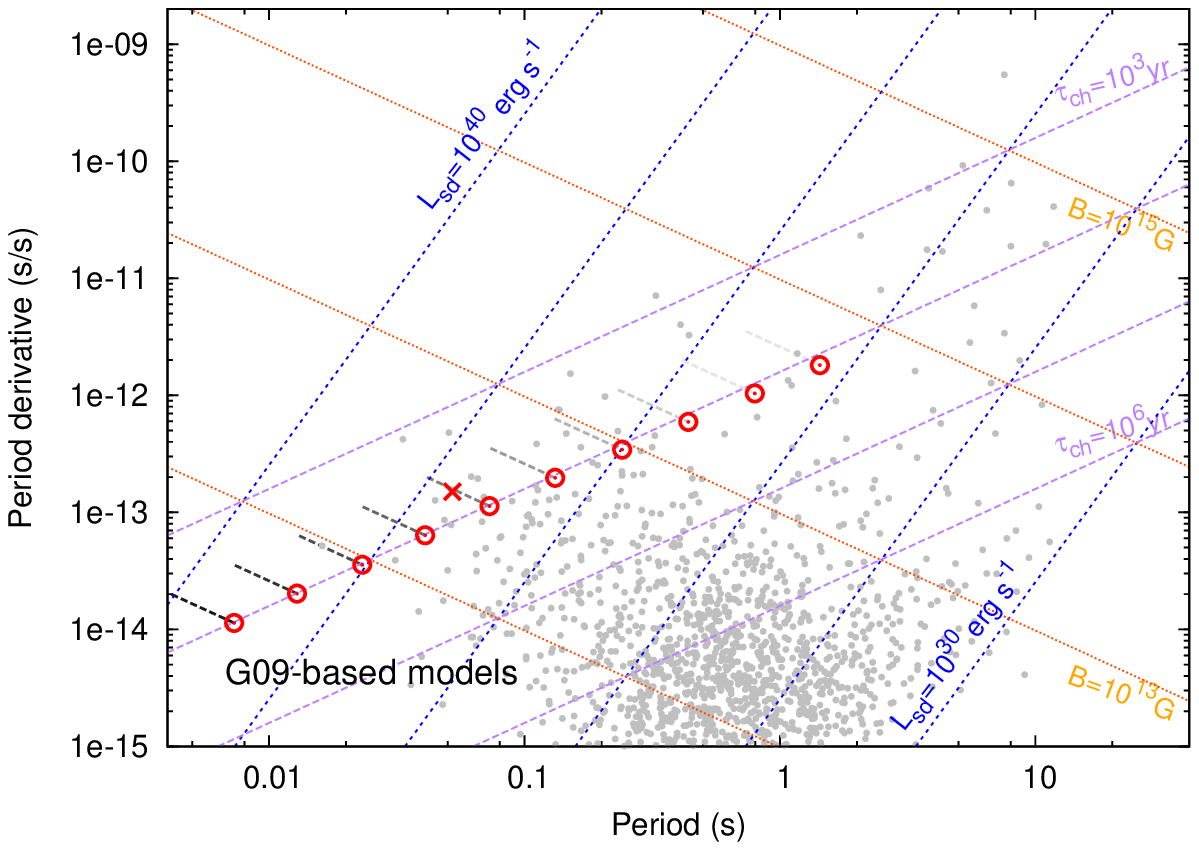}
                \includegraphics[width=0.45\textwidth]{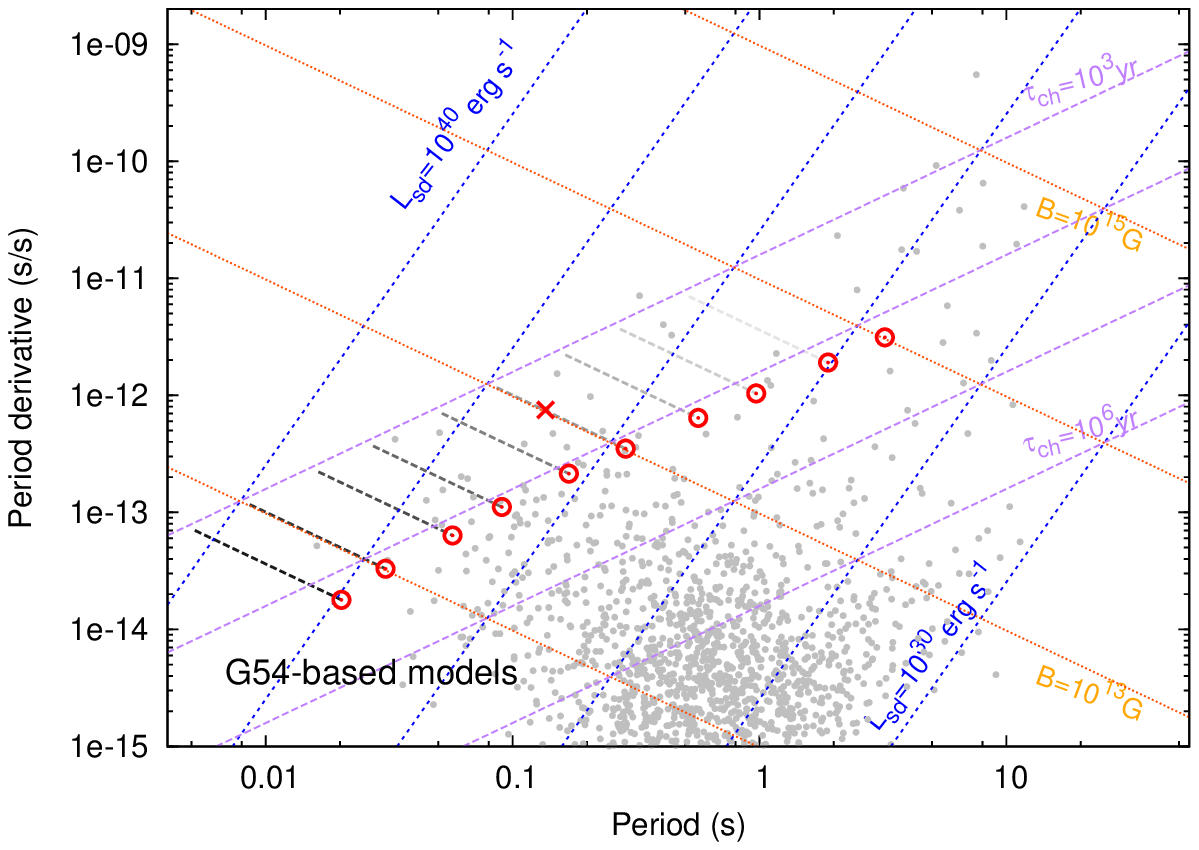}
    \caption{
    $P\dot{P}$-diagram for pulsars models using different anchor PWNe, as noted. The blue dashed lines represent an equal value of spin-down power, shown in increasing factors of $10^2$. The orange  and purple lines represent equal values of magnetic field and characteristic age, respectively, both are shown in increasing factors of 10. The grey points are known radio pulsars, obtained from the ATNF Catalog \protect\cite{Manchester2005}. The red crosses represent the current values of the corresponding anchor nebulae,
    whereas the hollow circles represent the values of $P$ and $\dot P$ at the moment of maximum X-ray efficiency for each of the models.
    The gray lines joining the red circles are the traces along their evolution. They are color-coded such that the darkest one represent pulsars with the highest initial spin-down power. 
    See text for further discussion.  }
    \label{PPdot}
\end{figure*}

We analyze 10 models for each of the anchor nebulae considered, and evolve each of them up to more than 20000 years of age in order to see how 
the whole reverberation process comes to an end and the continuing evolution thereafter.
{  Thus, each observed PWNe is assumed to generate a family of models obtained by changing their initial spin-down power. These families are explicitly shown in Fig. \protect\ref{PPdot}. The gridding used for the family conformation is admittedly coarse, but as can be seen in the figure (see the separation of the red dots and the corresponding evolutionary tracks) it already covers quite well the space of interest.}

\begin{figure*}
    \centering
    \includegraphics[width=0.49\textwidth]{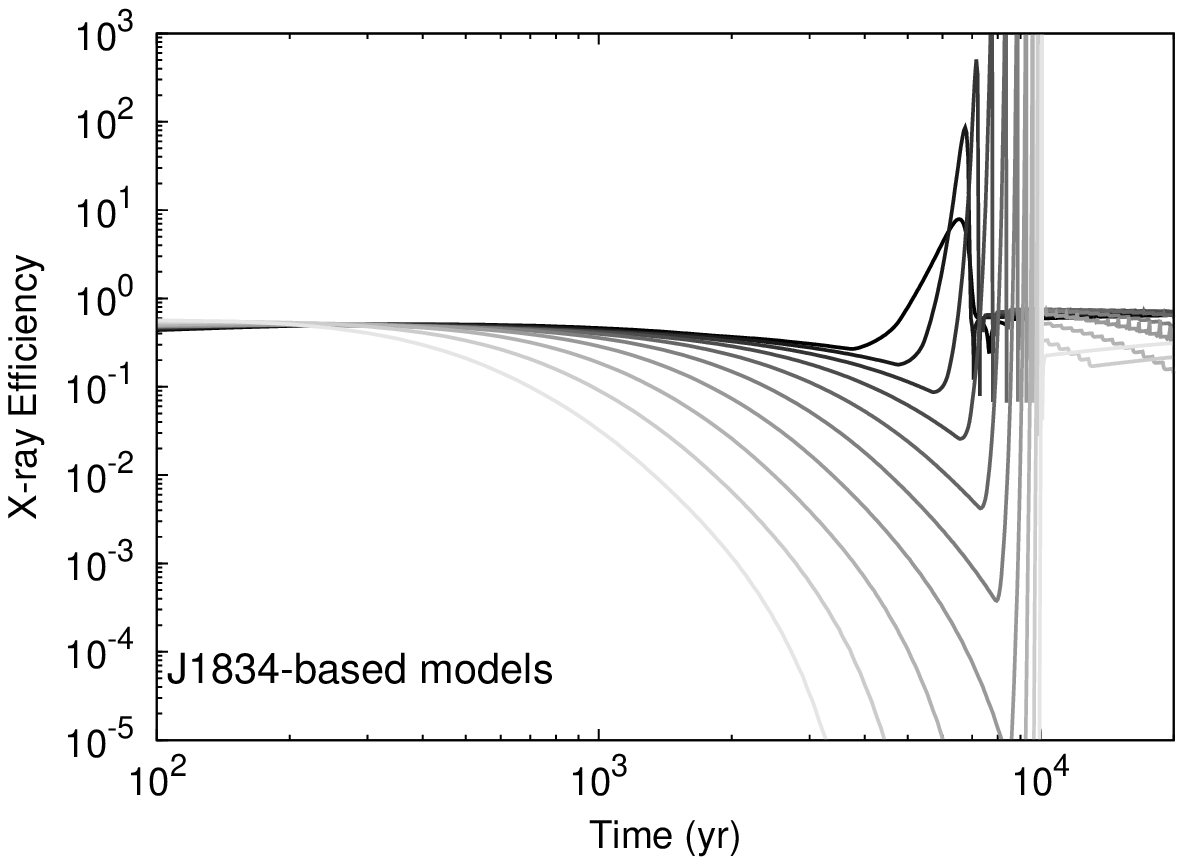}
    \includegraphics[width=0.49\textwidth]{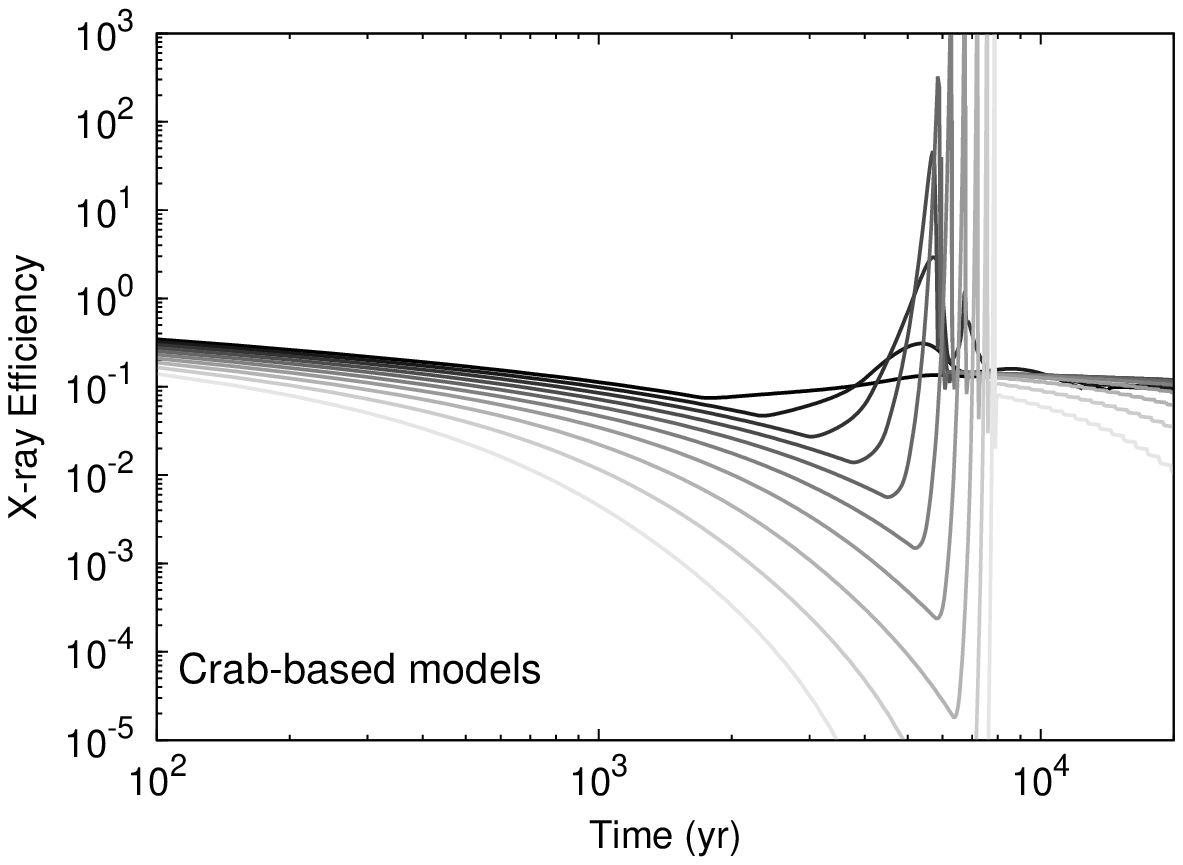}
    \includegraphics[width=0.49\textwidth]{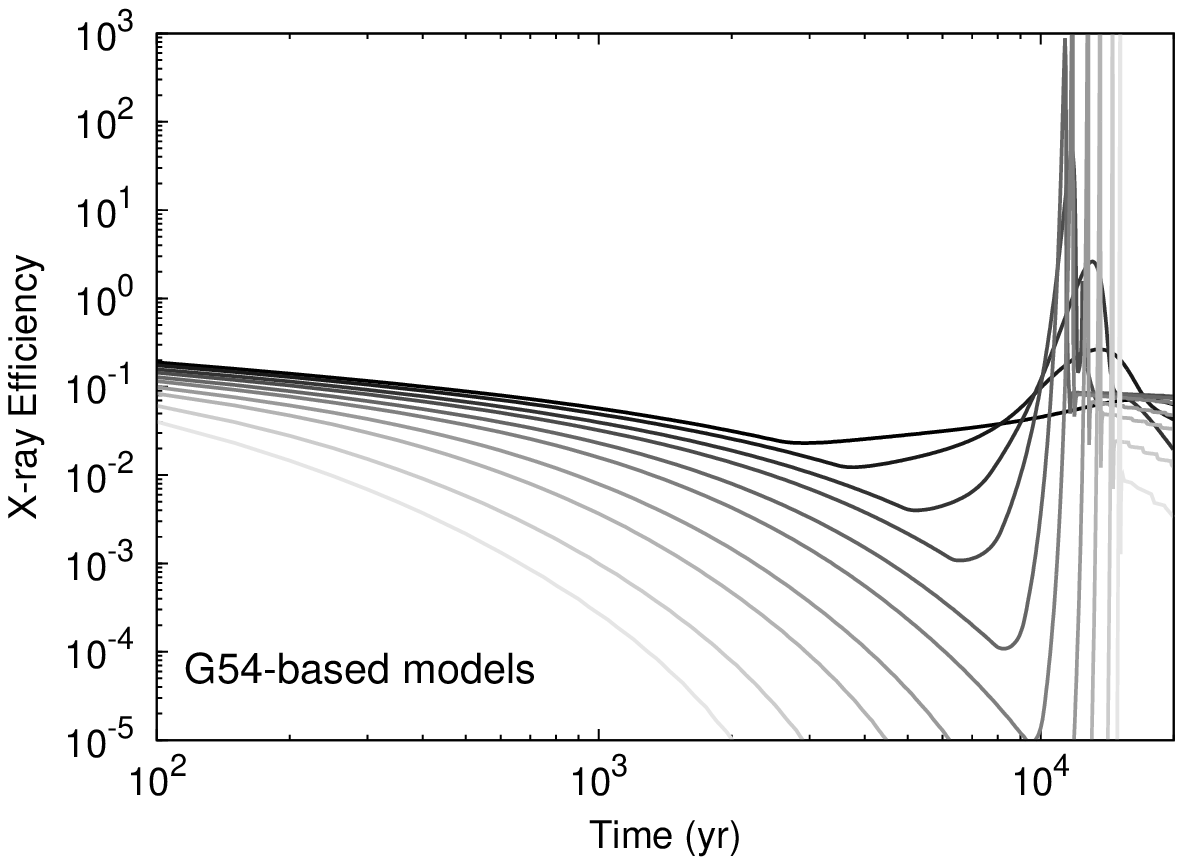}
    \includegraphics[width=0.49\textwidth]{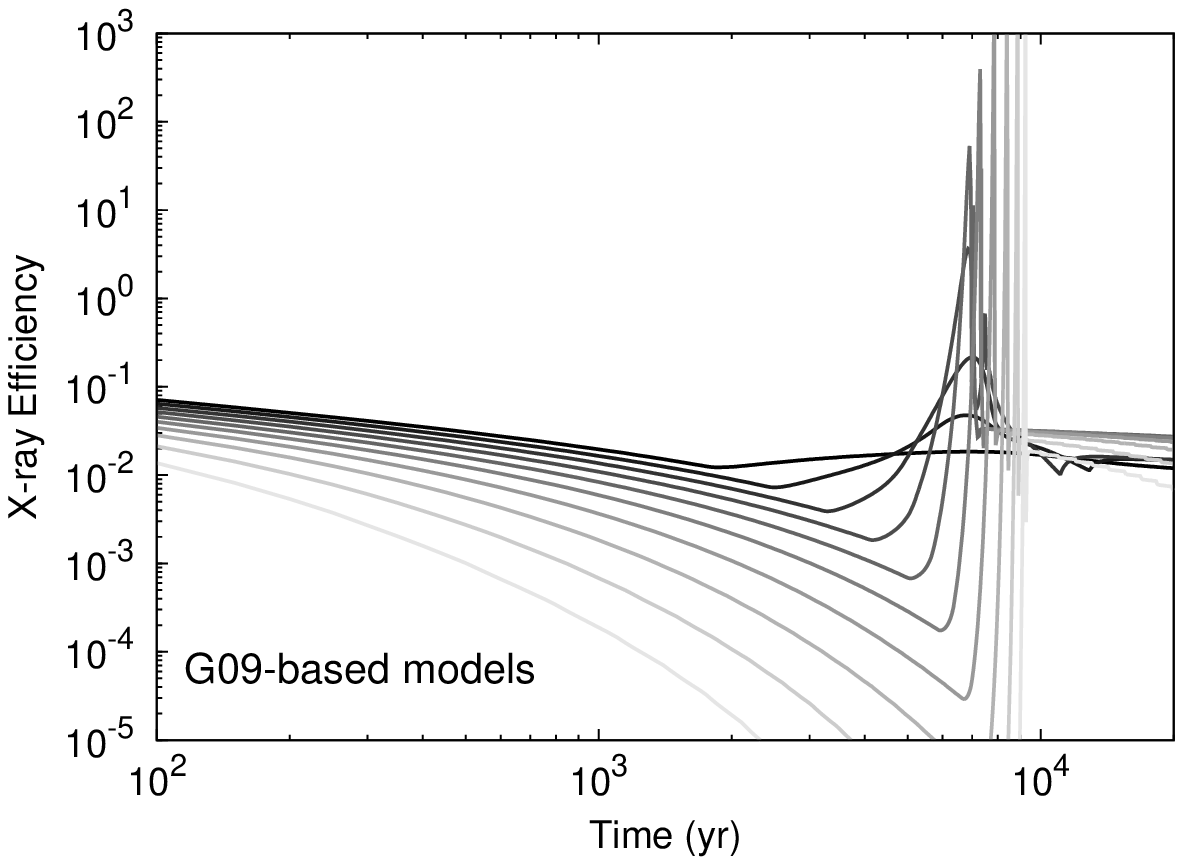}
 \caption{The evolution of X-ray efficiency for all models studied. The four panels correspond to each of the four anchor-PWNe analyzed. In each panel, the different grey levels are used to show the 10 corresponding models, with the lightest color representing the model with the lowest $L_0$, as in Fig. \protect\ref{PPdot}.
}
    \label{results0}
\end{figure*}

Table 1 also gives the maximal X-ray efficiency achieved, the
PWN age at which the X-ray maximal efficiency happens, the duration of the corresponding superefficient X-ray period as well as the starting and end of this period, (Dos$_{\rm X}$, $t_{\rm start-X}$, $t_{\rm end-X}$),
and the same for the whole reverberation process (Dor, $t_{\rm start-R}$, $t_{\rm end-R}$).
Models noted in red in Table 1 represent the values of the real, and corresponding anchor PWNe.
Given that we are not interested in fitting any particular observational data, we shall not present here the evolving SEDs or the electron spectrum for the simulated PWNe, but report 
on the properties of the reverberation period only.
Fig. \ref{results0} gives the evolution of the X-ray efficiency 
(in the 0.1--10 keV band) as a function of time for the models considered. 
We use these evolutionary plots next to determine the reverberation and superefficiency properties of these PWNe.

\begin{table*}
\centering
\caption{Models of PWNe. See text for a discussion.}
\begin{tabular}{lllrrrrrr}
\hline
\hline
 anchor PWN:                & &  & J1834.9-0846 & Crab & G0.9+0.1& G54.1+0.3 \\
 \hline

model 1   & $L_0$                   & [erg s$^{-1}$]              & $1.7\times 10^{40}$ &   $3.0\times 10^{40}$ &   $1.1\times 10^{40}$ &   $2.0\times 10^{40}$ \\ \hline
                & Max. Eff.$_X$		& \ldots		& 7.9		& 0.4 		& 0.02 	& 0.07 \\
                & $t$(max.eff)$_X$ & [years]     & 6550.0		& 5850.0 	& 6900.0 	& 16800.0 \\
                & Dos$_{X}$ & [years]   			& 1909.0		& 0 			& 0		& 0 \\
                & $t_{\rm start-X}$ & [years]	& 5141.0		& 0			& 0 		& 0 \\
                & $t_{\rm end-X}$ & [years]		& 7050.0 	& 0			& 0 		& 0 \\
                & Dor& [years]   		& 3282.3		& 5879.0 	& 0 		& 11180.0 \\
                & $t_{\rm start-R}$ & [years]	& 4370.0		& 3788.0 	& 0 		& 9991.0 \\
                & $t_{\rm end-R}$ & [years]		& 7652.3		& 9667.0 	& 0 		& 21171.0 \\
 \hline
model 2   & $L_0$ & [erg s$^{-1}$]              &   $5.5\times 10^{39}$  &   $9.5\times 10^{39}$  &   $3.5\times 10^{39}$  &   $7.2\times 10^{39}$  \\  \hline
                & Max. Eff.$_X$		& \ldots		& 85.8		& 0.3 		& 0.05 		& 0.3 \\
                & $t$(max.eff)$_X$ & [years]     & 6750.0		& 5400.0 	& 6800.0 	& 13550.0 \\
                & Dos$_{X}$ & [years]   			& 1146.0		& 0 			& 0 			& 0 \\
                & $t_{\rm start-X}$ & [years]	& 5804.0		& 0 			& 0 			& 0 \\
                & $t_{\rm end-X}$ & [years]      & 6950.0		& 0 			& 0 			& 0 \\
                & Dor& [years]		& 1963.3		& 5303.0 	& 4619.0 	& 10243.0 \\
                & $t_{\rm start-R}$ & [years]	& 5092.0		& 3423.0 	& 4293.0 	& 6878.0 \\
                & $t_{\rm end-R}$ & [years]		& 7055.3		& 8726.0 	& 8912.0 	& 17121.0 \\
 \hline
model 3   & $L_0$ & [erg s$^{-1}$]              & $1.7\times 10^{39}$      &   {\color{red} $3.0\times 10^{39}$}    &   $1.1\times 10^{39}$    &   $2.0\times 10^{39}$   \\  \hline
                & Max. Eff.$_X$		& \ldots		& 507.7		& 2.9 		& 0.2 	& 2.6 \\
                & $t$(max.eff)$_X$ & [years]     & 7150.0		& 5700.0 	& 7000.0 	& 13100.0 \\
                & Dos$_{X}$ & [years]   			& 674.3		& 907.6 		& 0 		& 2103.0 \\
                & $t_{\rm start-X}$ & [years]    & 6566.3		& 5142.4 	& 0 		& 11797.0 \\
                & $t_{\rm end-X}$ & [years]      & 7240.6		& 6050.0 	& 0 		& 13900.0 \\
                & Dor& [years]   		& 1400.4		& 3200.7 	& 4208.0 	& 8386.0 \\
                & $t_{\rm start-R}$ & [years]    & 5880.0		& 3641.0 	& 4236.0 	& 6680.0 \\
                & $t_{\rm end-R}$ & [years]      & 7280.4		& 6841.7 	& 8444.0 	& 15066.0 \\
 \hline
model 4   & $L_0$ & [erg s$^{-1}$]              &  $5.5\times 10^{38}$   &   $9.5\times 10^{38}$  &   $3.5\times 10^{38}$  &   $7.2\times 10^{38}$  \\  \hline
                & Max. Eff.$_X$		& \ldots		& 1917.2		& 46.3 		& 3.7 	& 45.3 \\
                & $t$(max.eff)$_X$ & [years]		& 7750.0		& 5700.0 	& 6850.0 	& 11750.0 \\
                & Dos$_{X}$ & [years]			& 450.8		& 690.6 		& 523.8 		& 1332.2 \\
                & $t_{\rm start-X}$ & [years]	& 7333.2		& 5159.4 	& 6526.2 	& 10717.8 \\
                & $t_{\rm end-X}$ & [years]		& 7784.0		& 5850.0 	& 7050.0 	& 12050.0 \\
                & Dor& [years]		& 1155.7		& 1857.5 	& 2910.4 	& 5193.0 \\
                & $t_{\rm start-R}$ & [years]	& 6644.0		& 4105.0 	& 4686.0 	& 7313.0 \\
                & $t_{\rm end-R}$ & [years]		& 7799.7		& 5962.5		& 7596.4 	& 12506.0 \\

 \hline
model 5   & $L_0$ & [erg s$^{-1}$]              & {\color{red} $1.7\times 10^{38}$}     &   $3.0\times 10^{38}$    &   {\color{red}$1.1\times 10^{38}$}    &   $2.0\times 10^{38}$    \\  \hline
                & Max. Eff.$_X$		& \ldots		& 5444.4		& 325.6 		& 54.2 		& 886.0 \\
                & $t$(max.eff)$_X$ & [years]		& 8336.5		& 5850.0 	& 6910.0 	& 11350.0 \\
                & Dos$_{X}$ & [years]			& 323.6		& 408.8 		& 360.0 		& 584.4 \\
                & $t_{\rm start-X}$ & [years]	& 8042.9		& 5521.5 	& 6616.6 	& 10846.1 \\
                & $t_{\rm end-X}$ & [years]		& 8366.5		& 5930.3 	& 6976.6 	& 11430.5 \\
                & Dor& [years]		& 1036.9		& 1289.3 	& 1702.9 	& 2960.9 \\
                & $t_{\rm start-R}$ & [years]	& 7337.0		& 4676.0 	& 5342.0 	& 8526.0 \\
                & $t_{\rm end-R}$ & [years]		& 8373.9		& 5965.3 	& 7044.9 	& 11486.9 \\

 \hline
model 6   & $L_0$ & [erg s$^{-1}$]              &  $5.5\times 10^{37}$    &   $9.5\times 10^{37}$  &   $3.5\times 10^{37}$  &   {\color{red}$7.2\times 10^{37}$}  \\  \hline
                & Max. Eff.$_X$		& \ldots		& 12507.8	& 1950.9 	& 393.3 		& 5464.0 \\
                & $t$(max.eff)$_X$ & [years]     & 8854.2		& 6272.1 	& 7297.3 	& 11812.3 \\
                & Dos$_{X}$ & [years]   			& 231.6		& 271.2		& 235.2 		& 348.4 \\
                & $t_{\rm start-X}$ & [years]    & 8640.6		& 6020.9 	& 7087.1 	& 11493.9 \\
                & $t_{\rm end-X}$ & [years]      & 8872.3		& 6292.1 	& 7322.3 	& 11842.3 \\
                & Dor& [years]   		& 931.4		& 1025.9 	& 1284.8 	& 2241.6 \\
                & $t_{\rm start-R}$ & [years]    & 7947.0		& 5281.0 	& 6059.0 	& 9620.0 \\
                & $t_{\rm end-R}$ & [years]      & 8878.4		& 6306.9 	& 7343.8 	& 11861.6 \\

 \hline

\end{tabular}
\end{table*}

\begin{table*}
\centering
\caption{Table 1, continued.}
\begin{tabular}{lllllll}
\hline
\hline
 anchor PWN:                & &  & J1834.9-0846 & Crab & G0.9+0.1& G54.1+0.3 \\
 \hline

model 7   & $L_0$ & [erg s$^{-1}$]              & $1.7\times 10^{37}$      &   $3.0\times 10^{37}$    &   $1.1\times 10^{37}$    &   $2.0\times 10^{37}$    \\  \hline
                & Max. Eff.$_X$		& \ldots		& 26368.0	& 7230.6 	& 1956.5 	& 24111.6 \\
                & $t$(max.eff)$_X$ & [years]     & 9270.0		& 6739.3 	& 7844.7 	& 12799.8 \\
                & Dos$_{X}$ & [years]   			& 157.6		& 198.9  	& 171.0 		& 217.7 \\
                & $t_{\rm start-X}$ & [years]	& 9123.3		& 6555.4 	& 7693.7 	& 12600.9 \\
                & $t_{\rm end-X}$ & [years]      & 9281.0		& 6754.3 	& 7864.7 	& 12818.6 \\
                & Dor& [years]		& 812.6		& 901.4 		& 1100.4 	& 1843.3 \\
                & $t_{\rm start-R}$ & [years]	& 8472.0		& 5861.0 	& 6774.0 	& 10984.0 \\
                & $t_{\rm end-R}$ & [years]		& 9284.6		& 6762.4 	& 7874.4 	& 12827.3 \\

 \hline
model 8   & $L_0$ & [erg s$^{-1}$]              &   $5.5\times 10^{36}$  &   $9.5\times 10^{36}$  &   $3.5\times 10^{36}$  &   $7.2\times 10^{36}$  \\  \hline
                & Max. Eff.$_X$		& \ldots		& 50724.6	& 17457.5 	& 6530.7 	& 57685.8 \\
                & $t$(max.eff)$_X$ & [years]		& 9599.4		& 7194.8 	& 8394.8 	& 13645.4 \\
                & Dos$_{X}$ & [years]			& 106.0		& 150.9 		& 142.7 		& 167.4 \\
                & $t_{\rm start-X}$ & [years]	& 9499.0		& 7052.4 	& 8263.8 	& 13487.4 \\
                & $t_{\rm end-X}$ & [years]		& 9604.9		& 7203.3 	& 8406.5 	& 13654.9 \\
                & Dor& [years]		& 688.8 		& 813.1 		& 987.3 		& 1667.2 \\
                & $t_{\rm start-R}$ & [years]	& 8918.0		& 6395.0 	& 7426.0		& 11993.0 \\
                & $t_{\rm end-R}$ & [years]		& 9606.8		& 7208.1 	& 8413.3 	& 13660.2 \\

 \hline
model 9   & $L_0$ & [erg s$^{-1}$]              &   $1.7\times 10^{36}$    &   $3.0\times 10^{36}$    &   $1.1\times 10^{36}$    &   $2.0\times 10^{36}$     \\  \hline
                & Max. Eff.$_X$		& \ldots		& 94499.4	& 40631.0 	& 18523.9 	& 145837.0 \\
                & $t$(max.eff)$_X$ & [years]		& 9859.3		& 7567.8 	& 8870.1 	& 14560.2 \\
                & Dos$_{X}$ & [years]			& 68.7		& 124.7 		& 116.0 		& 115.7 \\
                & $t_{\rm start-X}$ & [years]	& 9794.2		& 7454.5 	& 8760.8 	& 14450.4 \\
                & $t_{\rm end-X}$ & [years]		& 9862.8		& 7579.1 	& 8876.8 	& 14566.2 \\
                & Dor& [years]		& 568.1		& 717.4 		& 871.8 		& 1454.3 \\
                & $t_{\rm start-R}$ & [years]	& 9296.0		& 6862.0 	& 8009.0 	& 13115.0 \\
                & $t_{\rm end-R}$ & [years]		& 9864.1		& 7579.4 	& 8880.8 	& 14569.3 \\

 \hline
model 10   & $L_0$ & [erg s$^{-1}$]              &   $5.5\times 10^{35}$    &   $9.5\times 10^{35}$    &   $3.5\times 10^{35}$    &   $7.2\times 10^{35}$     \\  \hline
                & Max. Eff.$_X$		& \ldots 	& 166724.6		& 96896.9 	& 50159.2 	& 293947.0 \\
                & $t$(max.eff)$_X$ & [years]		& 10070.4		& 7871.0 	& 9241.3 	& 15141.3 \\
                & Dos$_{X}$ & [years]			& 62.7			& 77.2 		& 93.4 		& 94.8 \\
                & $t_{\rm start-X}$ & [years]	& 10010.0		& 7800.1 	& 9153.2 	& 15051.6 \\
                & $t_{\rm end-X}$ & [years]		& 10072.7		& 7877.2 	& 9246.6 	& 15146.3 \\
                & Dor& [years]		& 459.0			& 608.4 		& 744.8 		& 1259.9 \\
                & $t_{\rm start-R}$ & [years]	& 9614.0			& 7269.0 	& 8504.0 	& 13888.0 \\
                & $t_{\rm end-R}$ & [years]		& 10073.0		& 7877.4 	& 9248.8 	& 15147.9 \\

                \hline

\end{tabular}
\end{table*}


\begin{figure*}
    \centering
    \includegraphics[width=0.43\textwidth]{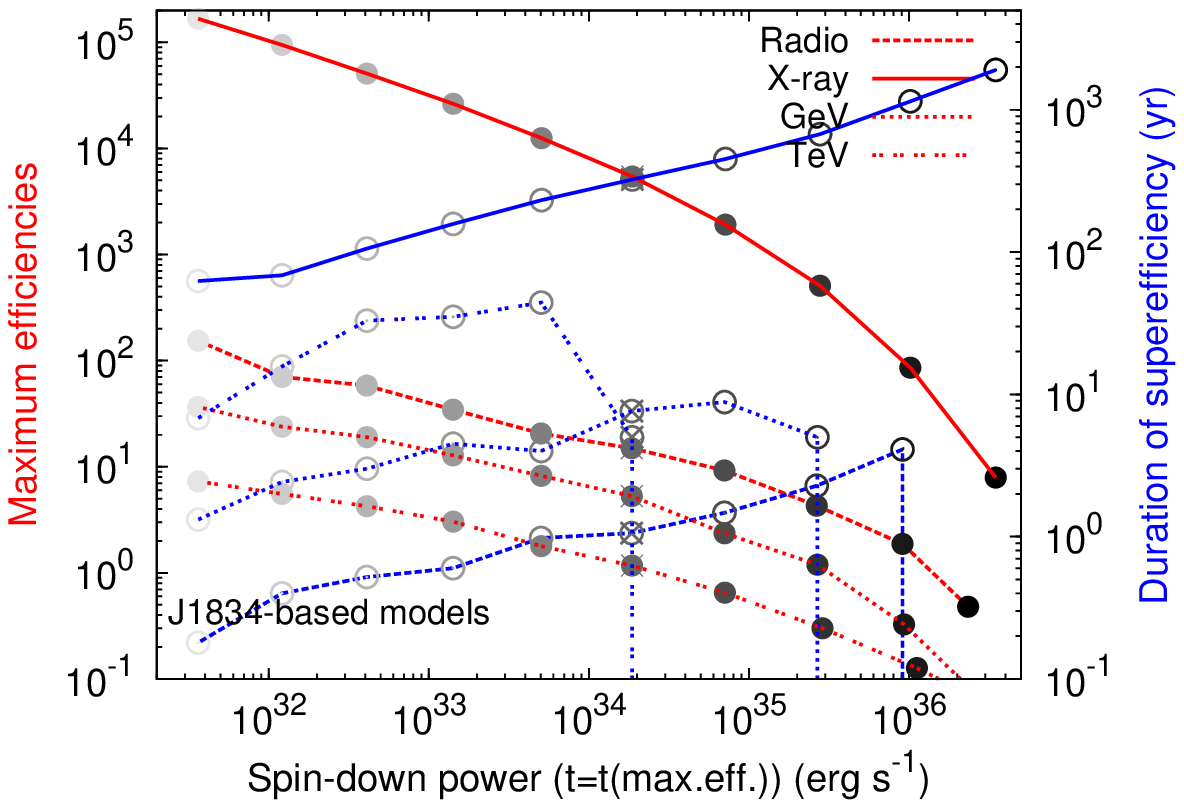}
    \includegraphics[width=0.43\textwidth]{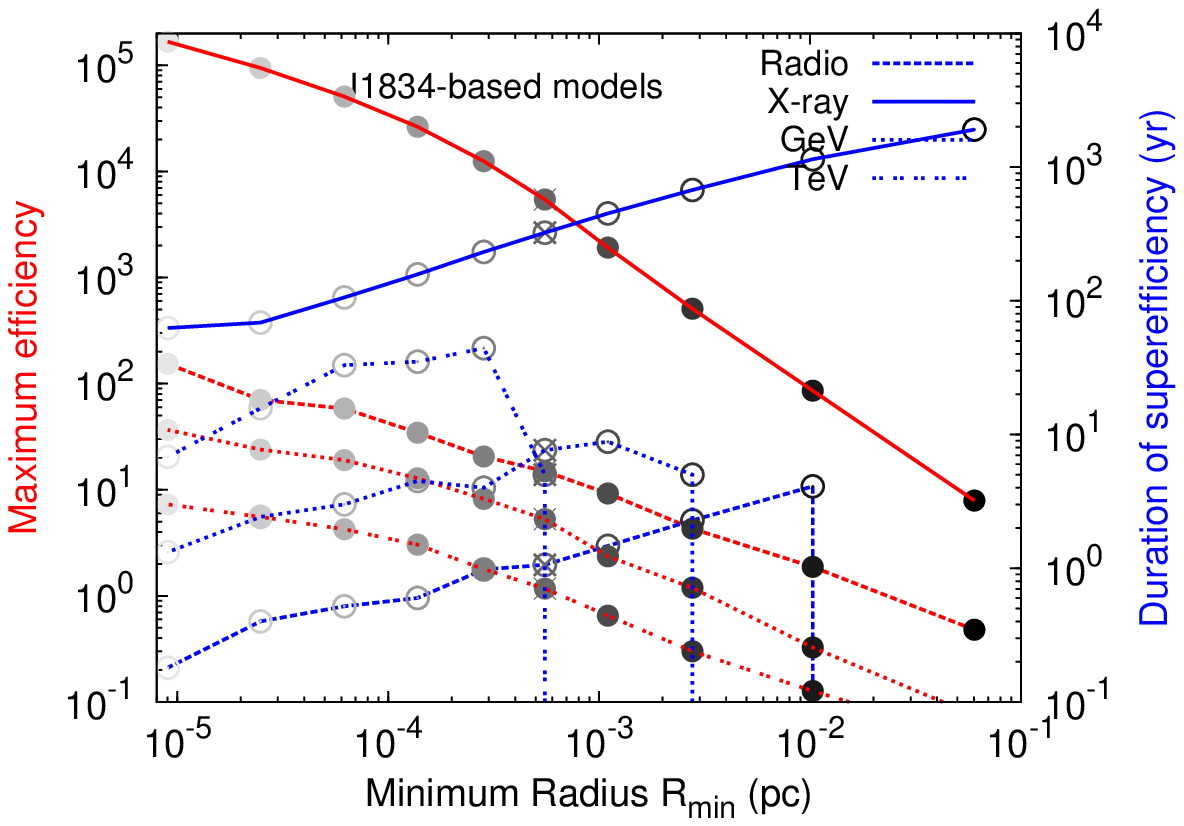}
    \includegraphics[width=0.43\textwidth]{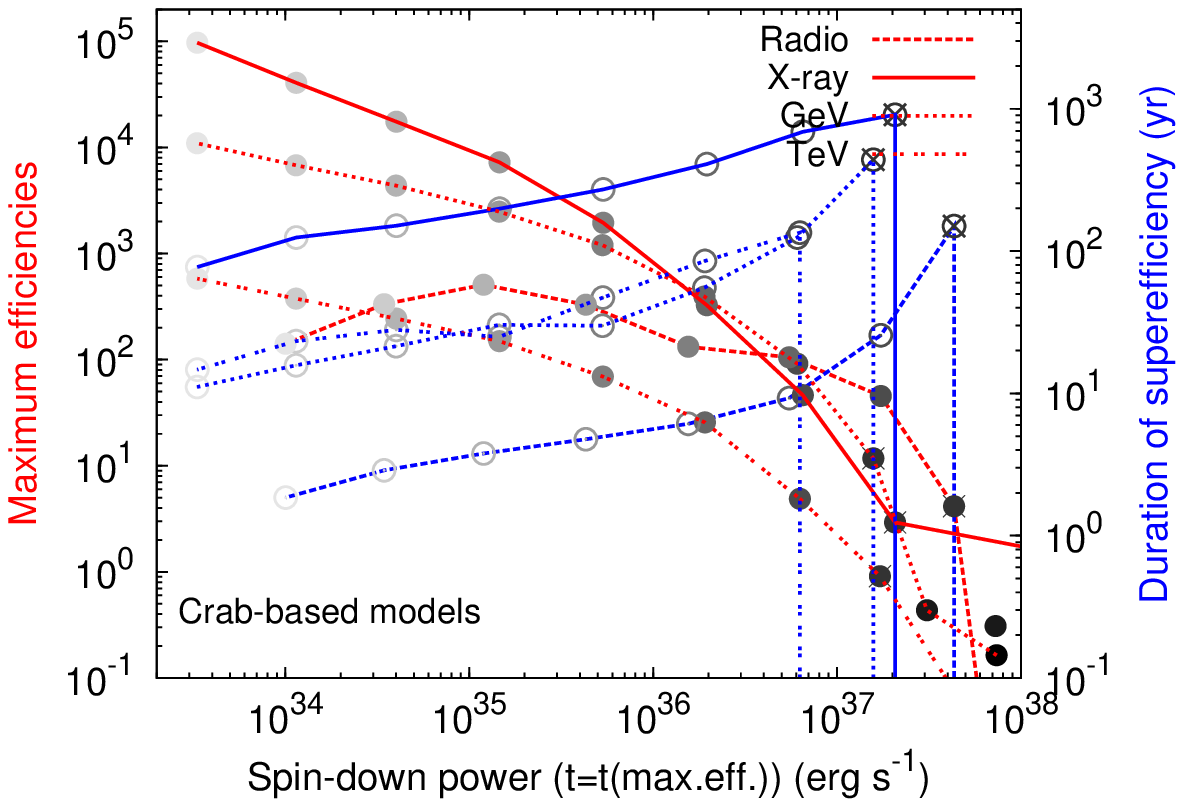}
    \includegraphics[width=0.43\textwidth]{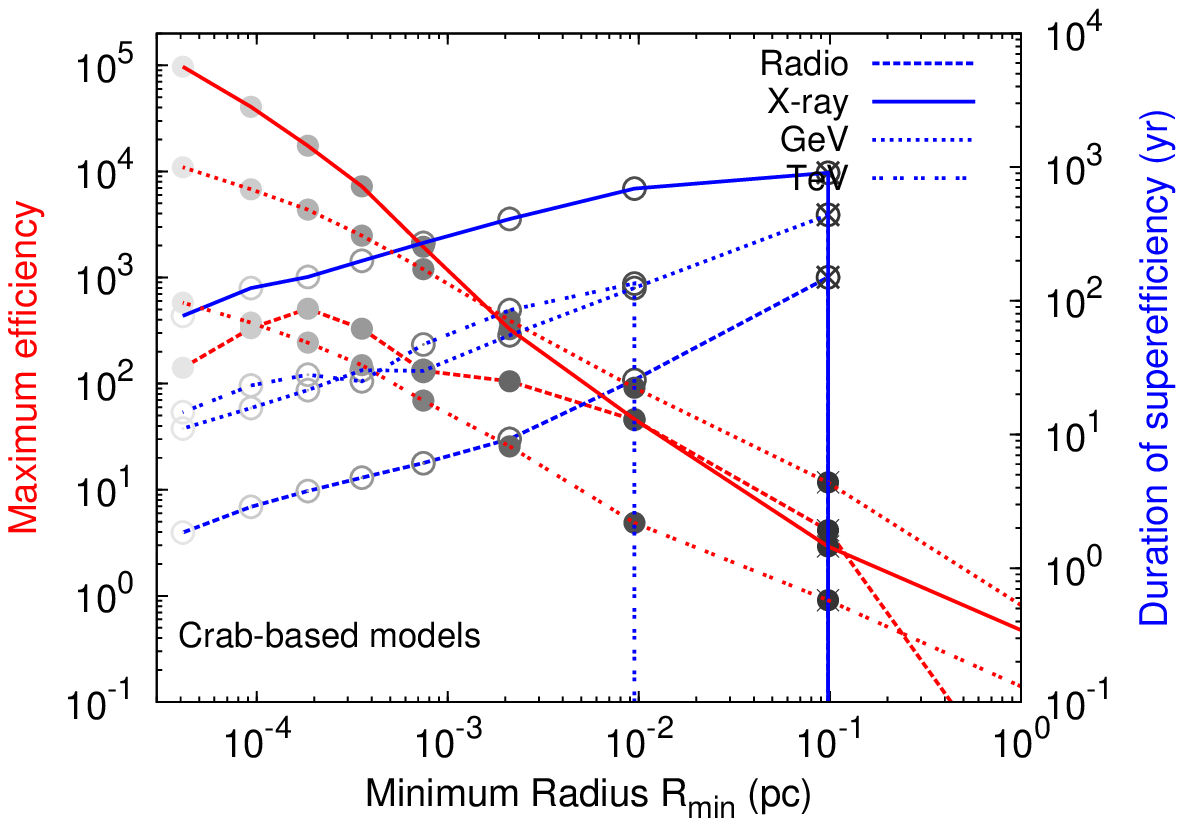}
    \includegraphics[width=0.43\textwidth]{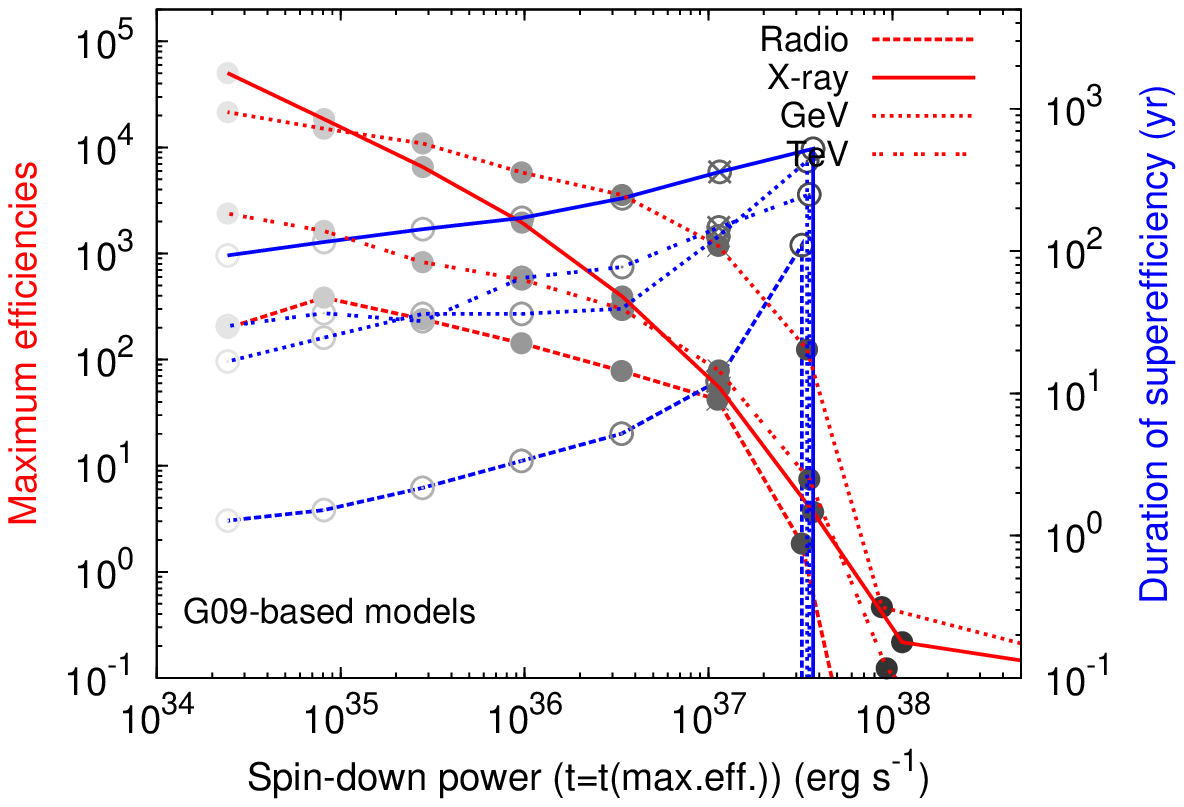}
    \includegraphics[width=0.43\textwidth]{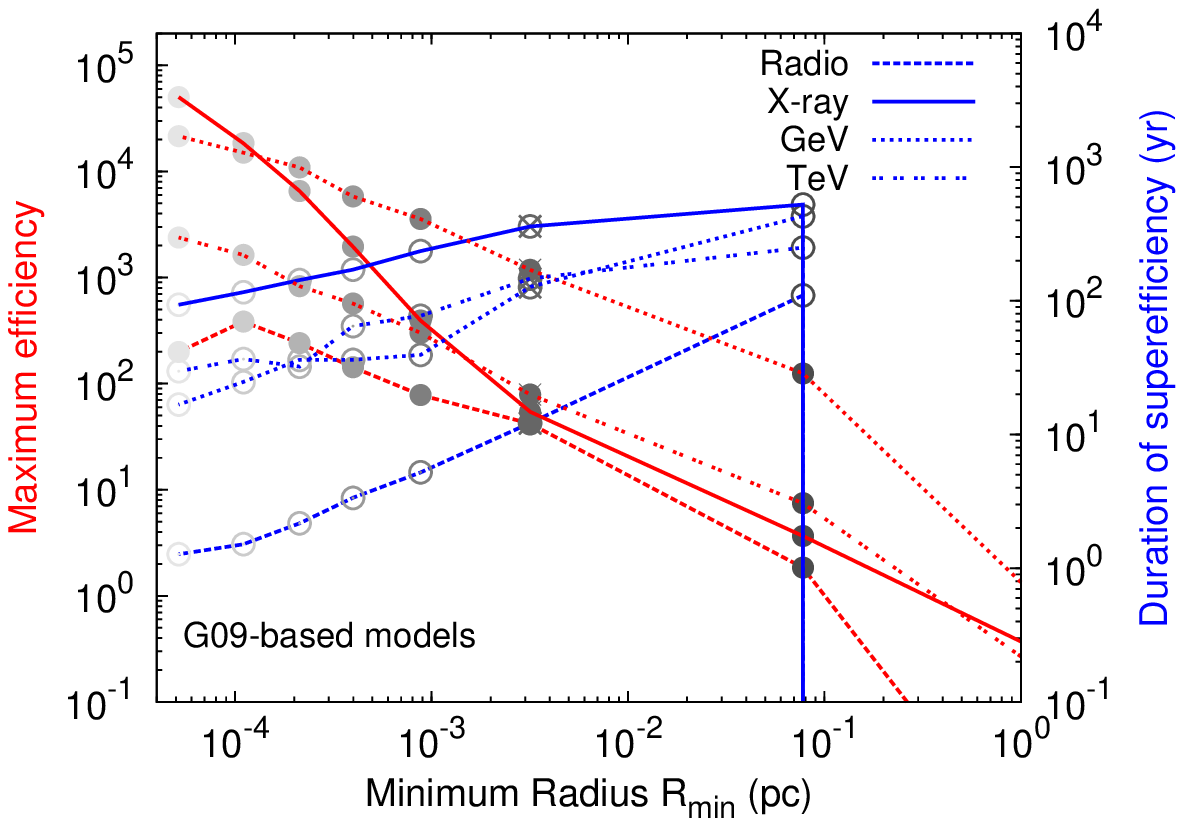}
    \includegraphics[width=0.43\textwidth]{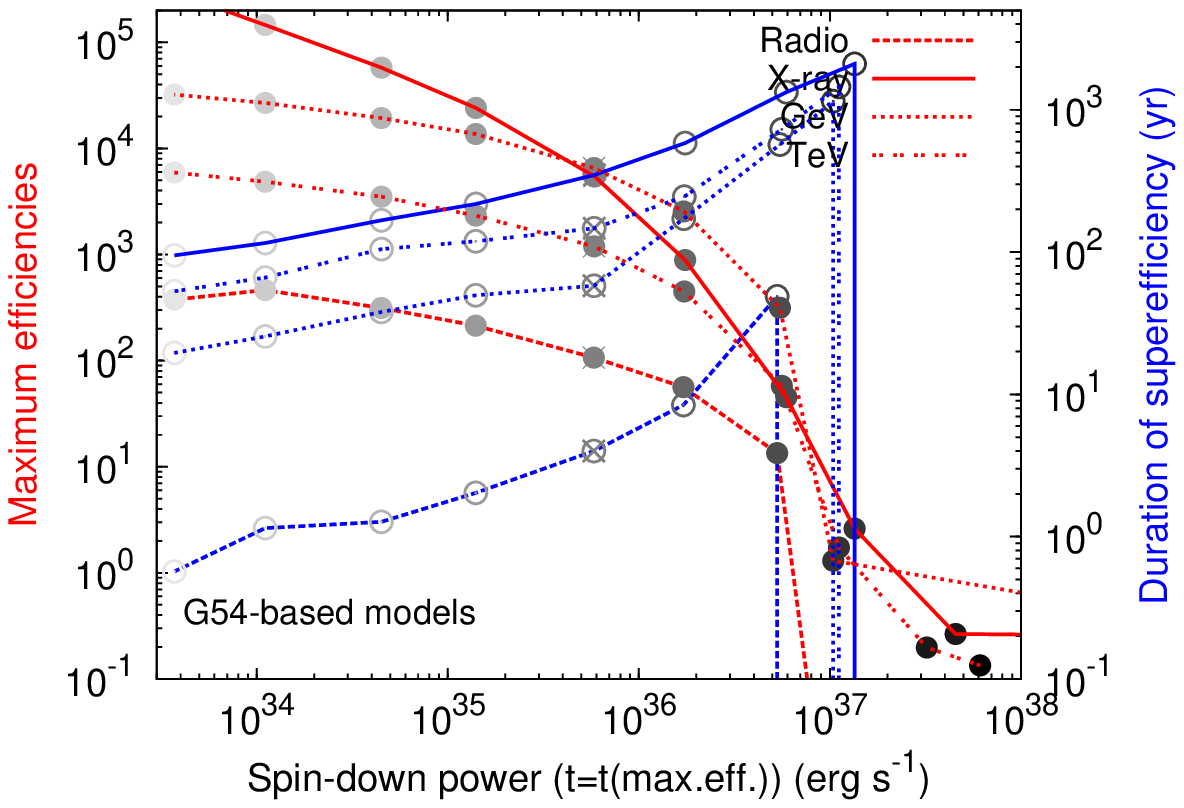}
    \includegraphics[width=0.43\textwidth]{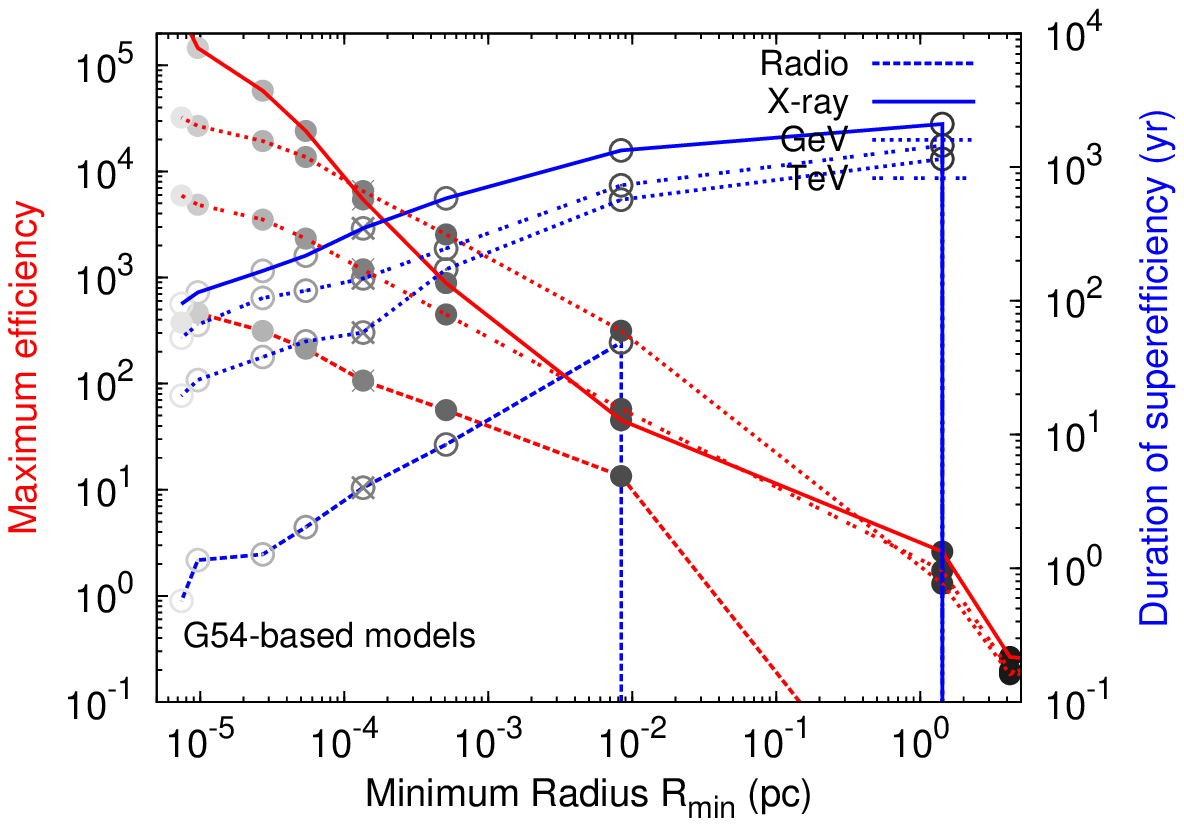}
    \caption{Exploring the superefficiency phase space.  These PWNe are similar in all respects (in particular, evolving in the same environment)  to each of the corresponding anchor-PWN quoted in the labels of the panels, but have varying initial spin-down power, see Table 1. 
The crosses along the curves of each panel show the properties of the superefficiency period  of the real PWN, as modelled in  \protect\cite{Torres2018b}. The dots along the curves represent the different models studied here, and are color-coded as in Fig. 1, with the lightest color assigned to the smallest initial spin-down power.
Maximum efficiencies are shown by the red lines, corresponding to the left vertical axis, while the duration of the superefficiency periods 
are shown by blue lines, referenced to the right vertical axis. Different type of lines denote different energy bands. 
}
    \label{expectations}
\end{figure*}

\subsection{Results}

The results for the superefficiency interval are summarized in Fig. \ref{expectations} and Table 1.  
The values of $t$(max.eff)$_X$ referred in Table 1 represent the time it takes (since its birth) 
to the putative PWN in reaching 
its maximum X-ray efficiency level. 
The latter is defined as the ratio between the X-ray luminosity and the spin-down power, which is  a function of time.
The  $t$(max.eff)$_X$ time interval   is also depicted in the corresponding Fig. \ref{PPdot} panel with a line representing the evolution since birth ($L_{sd}=L_0$) 
up to that moment. Whether that maximum
efficiency happens `today' (i.e., approximately at the time the dots representing ATNF pulsars in Fig. \ref{PPdot} are plotted) 
depends of course on the real time of birth of such putative pulsars.

The panels of Fig. \ref{expectations} show the maximum efficiency in different bands and the corresponding duration of the superefficiency interval as a function of the spin-down power at the time at which the corresponding maximum efficiency is reached (left), and of the minimum radius that is attained (right). 
Each panel refers to a given set of PWNe constructed using the same anchor system, as referred to in the plot. As described above, models within each panel 
all are subject to the same environment.
Also, for a given pulsar, the spin-down power is assumed to be uniformly decreasing with time as per the set of equations above, and all sets of models assume  
the same approximate range of initial spin-down powers as given in Table 1.
This is color-coded in the dots representing the models within Fig. \ref{expectations}, with the lightest color assigned to the smallest initial spin-down power.

Within the same panel, the maximum efficiencies happen at different ages of the PWN, because the underlying models have different energetics,
and thus the studied PWNe have different rotational power
when the nebula reaches the maximum efficiency.
From one set of models (one panel) to another, the assumed pulsar properties  and 
environment also change, contributing to additional differences due to their different pulsar evolution, e.g., because of having a different braking index and being subject 
to a different medium feedback.
Models with larger braking index and smaller age have larger spin-down power at $t=t$(max.eff).

The left panels of Fig.  \ref{expectations} show that for PWNe evolving in a similar environment, the maximum efficiencies are uniformly decreasing with the spin-down power,
whereas the duration of the process is longer.
When the central energy is larger and the pulsar is more powerful, there is a larger pressure in the PWNe  they generate, 
better supporting the environmental feedback. 
As a result, the compression is smaller and lasts a longer time.
Radio, GeV and TeV superefficiency may appear as well, but not for sufficiently energetic pulsars. 
For instance, for J1834-based models, there is no superefficiency at high energies as soon as the pulsars' spin-down power at the time of the maximum exceeds $\sim 2 \times 10^{35}$
and  $\sim 2 \times 10^{34}$ erg s$^{-1}$ for superefficiency at the GeV and TeV bands, respectively.
Equivalently, there is no superefficiency at the GeV and TeV bands
as soon as $L_0$ exceeds $\sim 5.5 \times 10^{37}$ erg s$^{-1}$ and $\sim 5.5 \times 10^{36}$ erg s$^{-1}$, respectively.
For the third set of models, based on G09,
superefficiency in all the  4 bands disappears when the 
spin-down power at $t=t$(max.eff) exceeds $\sim 2 \times 10^{37}$ erg s$^{-1}$, or when $L_0$ exceeds $\sim 3.5 \times 10^{37}$ erg s$^{-1}$.

The right panels  of Fig.  \ref{expectations} show that the minimum radius attained is (correspondingly) smaller for the more efficient nebulae.
The minimum radius  can reach in extreme models values as small as just a few times $10^{-5}$ pc, albeit this is not the common case.
However, although small, and as we noted earlier, even the more extreme cases produce PWNe that are
orders of magnitude larger than the pulsar's magnetosphere (typically at least 5 orders of magnitude larger than the size of a young pulsar's light cylinder), and  pulsed emission via synchro-curvature radiation (e.g., see  \cite{Torres2018}) is not expected to be affected.

The crosses in all curves of each panel of Fig. \ref{expectations} 
show the properties of the superefficiency period  of the real PWN.
When, for instance, J1834 reaches the maximum X-ray efficiency, the spin-down power 
will be of the order of $2 \times 10^{34}$ erg s$^{-1}$, not far from its current value, being the X-ray maximum just a few hundreds years ahead.
There are small differences as to when the maximum at other energy ranges happens in comparison with that in X-rays, what is reflected -albeit difficult to see in this scale at times- in 
Fig.~ \ref{expectations}.

Both the maximum efficiency and duration of the X-ray and radio superefficiency period (and every other frequency in between) are typically larger than those corresponding to higher-energy bands, what is especially notorious in the J1834-based models. 
The physical reasons why this happens were discussed in detail in \cite{Torres2018b}, see  figure 3 and the associated discussion there. 
From this perspective, then, searches in radio, optical or X-rays are likely to be more relevant than in gamma-rays.
Additionally, given that the angular resolution of lower-frequency instruments is better in comparison with gamma-ray ones, it seems that these frequencies are naturally preferred for a search.
Despite this fact, higher-energy instruments able to do unbiased surveys (e.g., using {\it Fermi}-LAT or the Galactic survey of H.E.S.S. or CTA) may uncover good candidates.

We note that less superefficient PWNe, with maximum X-ray efficiencies in the range of a few hundreds, or a few tens, 
will have a larger probability to be found in comparison to PWNe having more extreme reverberation episodes,
given that they spend a longer fraction of their life in this period.
For pulsars of small spin-down power (from $ 10^{32}$ to a few times $ 10^{34}$ erg s$^{-1}$) the X-ray maximum efficiency 
can be at least a factor of 1000 larger than the spin-down power at the time. 
For instance, taking the models based on J1834 as example, the model 9 has one of the largest X-ray efficiency of all, reaching $\sim 9 \times 10^4$, which implies a luminosity 
$L_X \sim 10^{37}$ erg s$^{-1}$ 
at that time. 
However, the superefficient period as a whole is active only for 68 years, and only for 16 years the X-ray efficiency is 
larger than $10^3$ ($L_X \sim 10^{35}$ erg s$^{-1}$). 
J1834 itself, with a current spin-down power 
$L_{sd} \sim 1.87 \times 10^{34}$ erg s$^{-1}$ 
and a future maximum superefficiency of 5444 will have a peak X-ray luminosity $L_X \sim 10^{38}$ erg s$^{-1}$. 
The period of X-ray superefficiency lasts in this case for 316 years, and the period for efficiency larger than 1000 would only be $\sim 60$ years.  
Less superefficient PWNe such as the J1834-based model 1 and others alike, with maximum X-ray efficiency of $\sim 8$, 
may have a larger probability to be found, given that the period of interest lasts 1900 years.
In any case, these are strong but short events in the life of stars, and will clearly be difficult to find.

\subsection{How many superefficient PWNe are there in the Galaxy?}

In our initial work, we roughly estimated that there could not be more than a few PWNe 
in an active superefficient stage today (or at any given time) in the Galaxy.
Given that the superefficiency period of the real PWNe studied in \citet{Torres2018b}
lasted a few hundreds years, we were interested in  $\sim $3\% of the evolution of the pulsars born in the last ten millennia.
Thus, given the pulsar birth rate, we would not expect more than a few Galactic nebulae to be currently superefficient.

\begin{figure}
    \centering
    \includegraphics[width=0.49\textwidth]{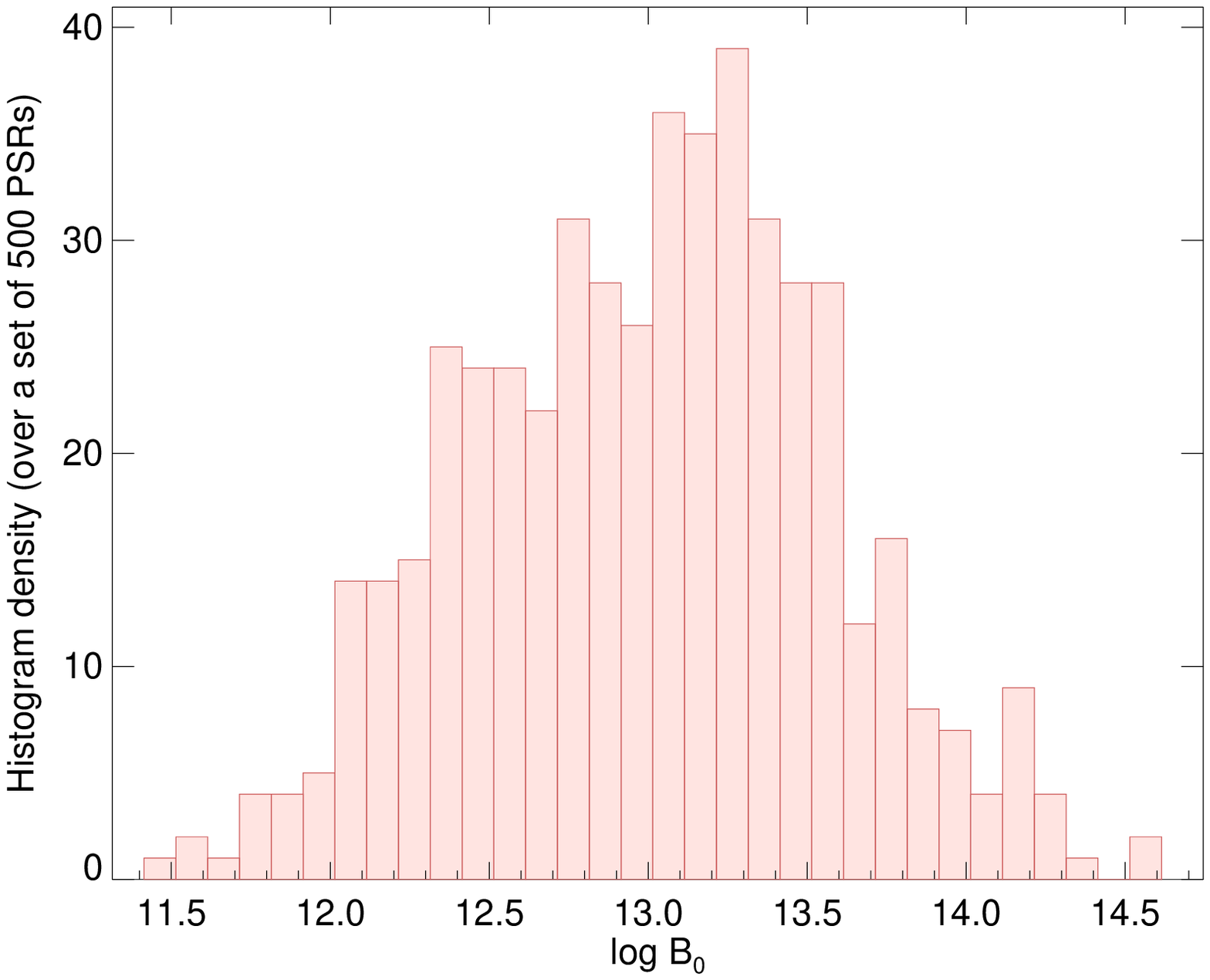}
      \includegraphics[width=0.49\textwidth]{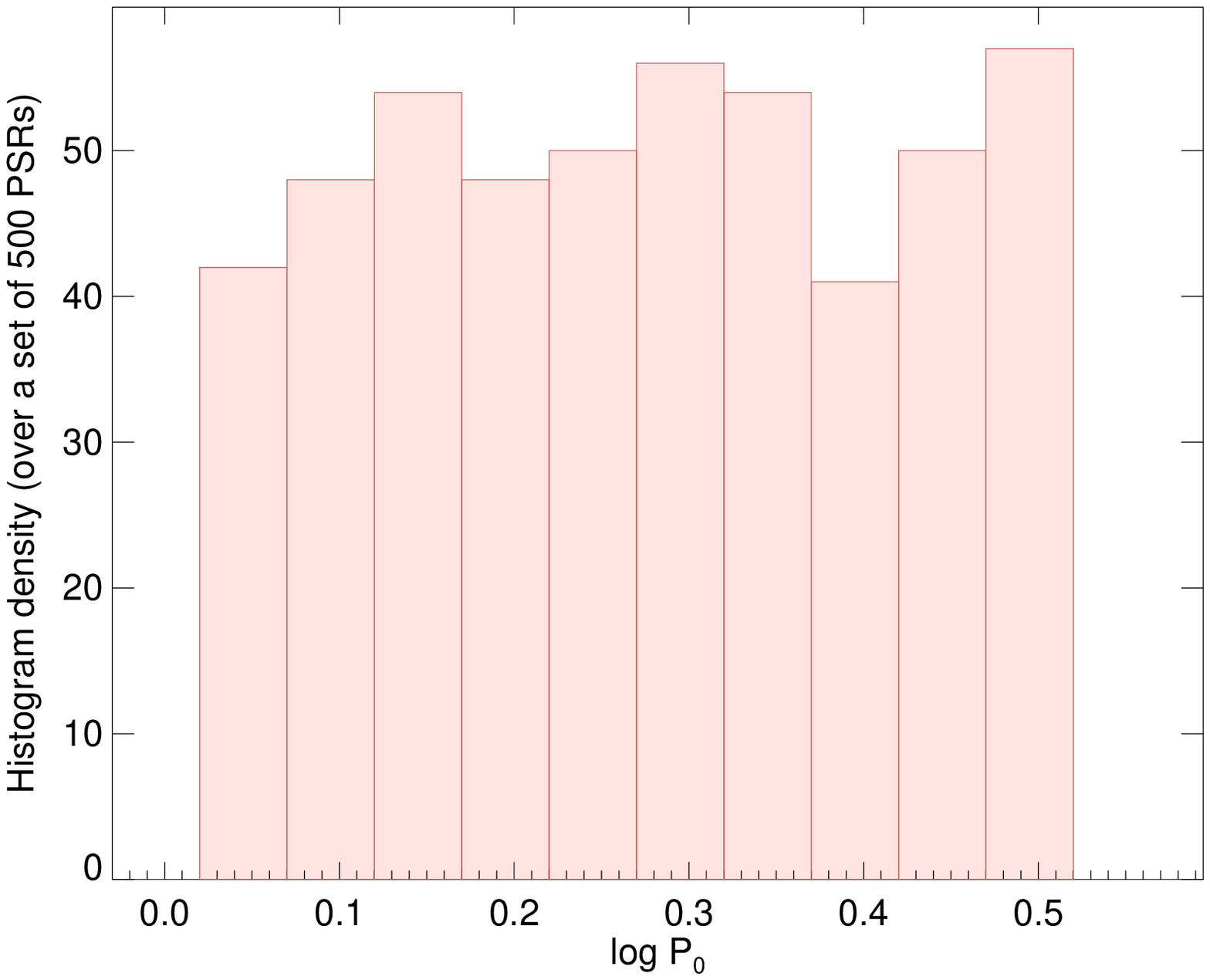}
    \includegraphics[width=0.49\textwidth]{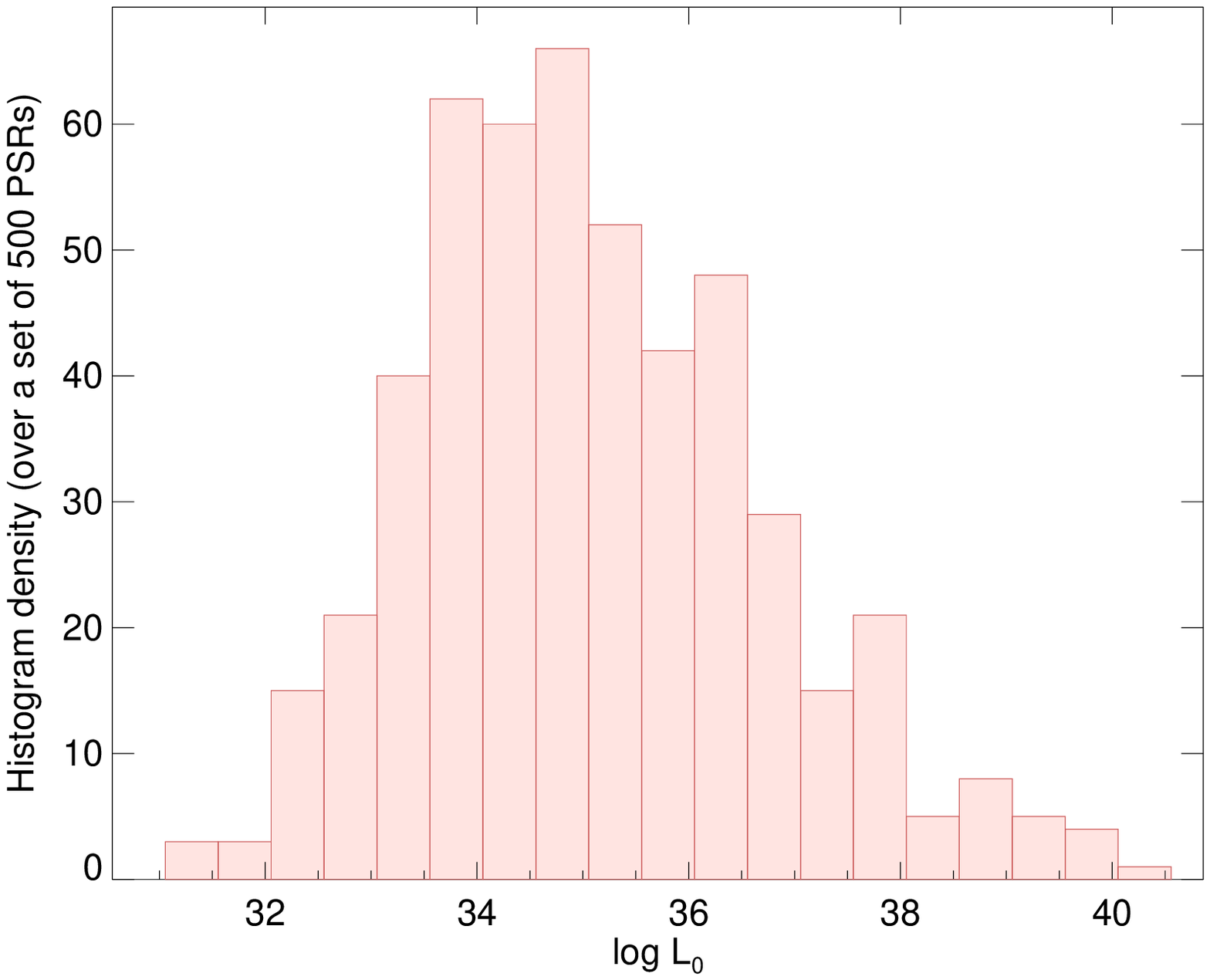}
    \caption{ Example of the random distributions of $B_0$, $P_0$, and $L_0$ considered for a set of 500 pulsars.}
    \label{dis}
\end{figure}

{  
In the framework of our specific radiative/dynamical model, one can obtain a more specific assessment. }
Since we now have many PWNe models studied in detail, and that for each of them we know when (with respect to the pulsar birth) and for how long their superefficiency stage happens, we can
improve the reliability of our earlier estimate.
To do so we shall assume the birthrate of pulsars in the Galaxy is 2.8 per century, as determined by  \citet{Faucher2006}.
We shall randomly associate to each member of a set of 500 pulsars (the number of pulsars appearing in the Galaxy in the last 17800 years, according to the birth rate)
a given value of initial magnetic field and period. 
We shall do so in a way that is compatible with the results
obtained in population synthesis
models of isolated neutron stars with magneto-rotational evolution \citep{Gullon2014}.
Thus, we shall consider that the logarithm of the initial magnetic field (in Gauss) with which pulsars are born is described by a Gaussian with mean of 
13 and a standard  deviation of 0.6; and that the initial period distribution is uniform for values less than 0.5 seconds. 
With these assumptions, and the fact that the 
magnetic field is described as usual by 
  $
  B(t) = 6.4 \times 10^{19} \sqrt { {P(t) \dot P(t)}/{{\rm s} }   }  \;
  {\rm G},
   $
{   (i.e., under the assumption of magnetic dipole domination of the Poynting flux, see \cite{Petri2015}) }
the distribution of initial spin-down powers is fixed using Eq. \ref{ePdot}; since
$L_{sd} \propto B^2 / P^4$.
Fig. \ref{dis} shows an example of the resulting distribution of initial magnetic fields, periods,
and spin-down powers for a simulated set of the youngest 500 pulsars born in the Galaxy that is compatible with population synthesis models.
In order to obtain averaged results, we repeat this random assignment of initial pulsar properties, generating 5000 different sets  (further runs do not change the results) 
of 500 pulsars, always agreeing with population synthesis models. 
For each pulsar in a simulated set, we thus know its age (the randomly assigned $t_{birth}$ according to the pulsar birth rate assumed) and its energetics.
If the initial spin-down power is within the range explored in Table 1, we consider the closest $L_0$-values among the cases studied there to represent the PWN evolution. If so, 
we know that the PWN will be in a superefficient stage today if 
$t_{\rm start-X} < t_{birth} < t_{\rm end-X}$, where $t_{\rm start-X} $ and $t_{\rm end-X}$ are the initial and final times for the X-ray superefficient period. Of course, 
$t_{\rm end-X}-t_{\rm start-X}={\rm Dos}_{\rm X}$. Specific values are given in Table 1.
If the initial spin-down power of the simulated pulsar  is off the range of simulated models in Table 1 (either from below or from above) we shall consider that no superefficiency happens.
More luminous pulsars continue the trend of decreasing maximum efficiency shown in Fig. \ref{expectations}, and indeed no superefficiency happens. These pulsars
are in any case unlikely to appear, being in the tail of the
distribution of $L_0$.
Less luminous pulsars than those in Table 1 may have superefficient events, but are so short in duration that in practice they also contribute little or nothing to the total expected number.

On average after 5000 simulated sets, we find that out of the last 500 pulsars 
born in our Galaxy, simulated according to population synthesis results, 
308.7 $\pm$ 11.0
of them are assigned an $L_0$ value within the range explored in Table 1.
And out of them, the mean number of superefficient  PWNe is 1.8 $\pm$ 1.3. The last panel of Fig. \ref{dis}
shows the histogram density for the number of superefficient PWNe out of the 5000 simulations.
At any given time in the Galaxy, then, our results imply that typically 
between 0 and 3 PWNe should be in a superefficient stage.

 \begin{figure}
    \centering
    \includegraphics[width=0.49\textwidth]{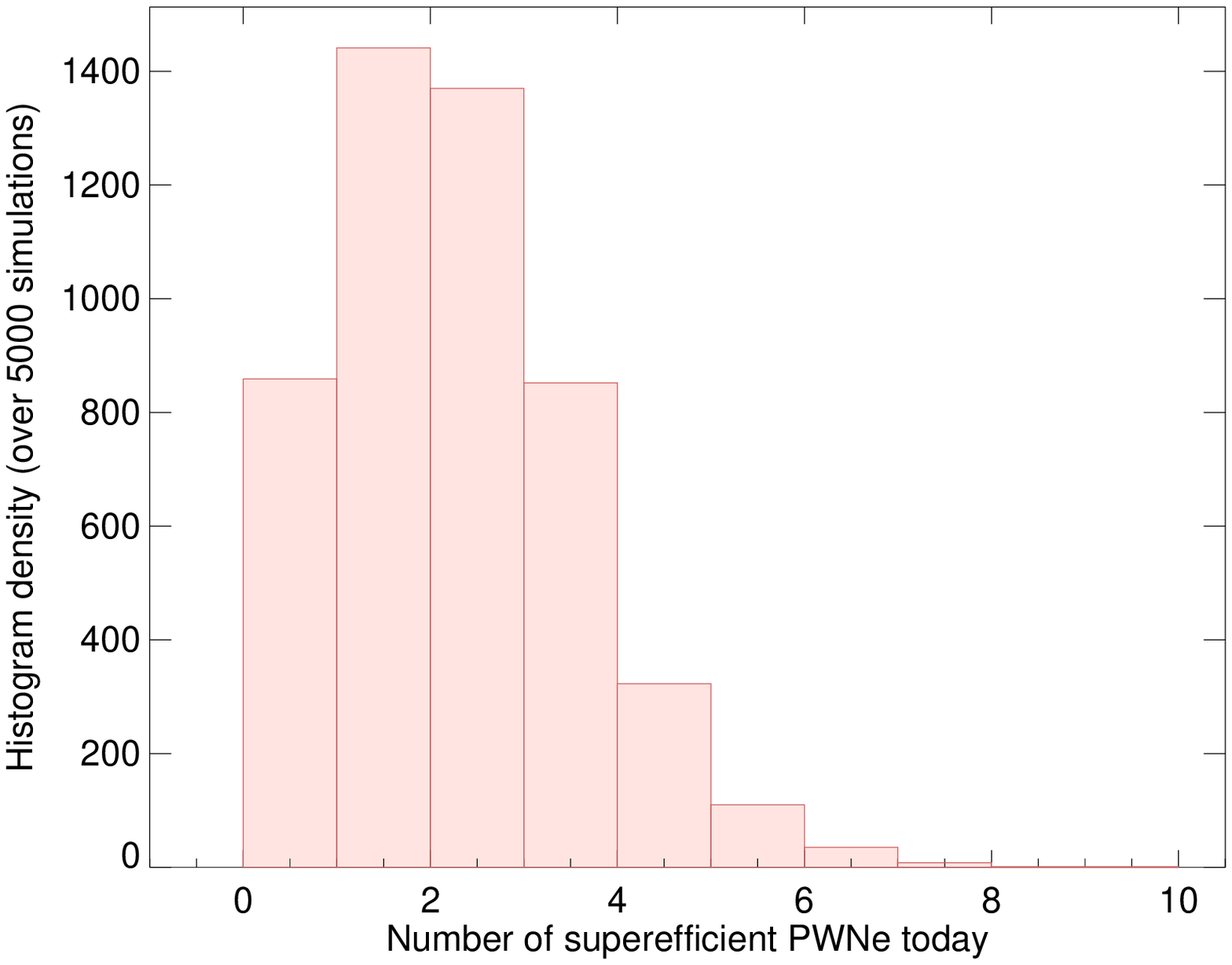}
        \includegraphics[width=0.49\textwidth]{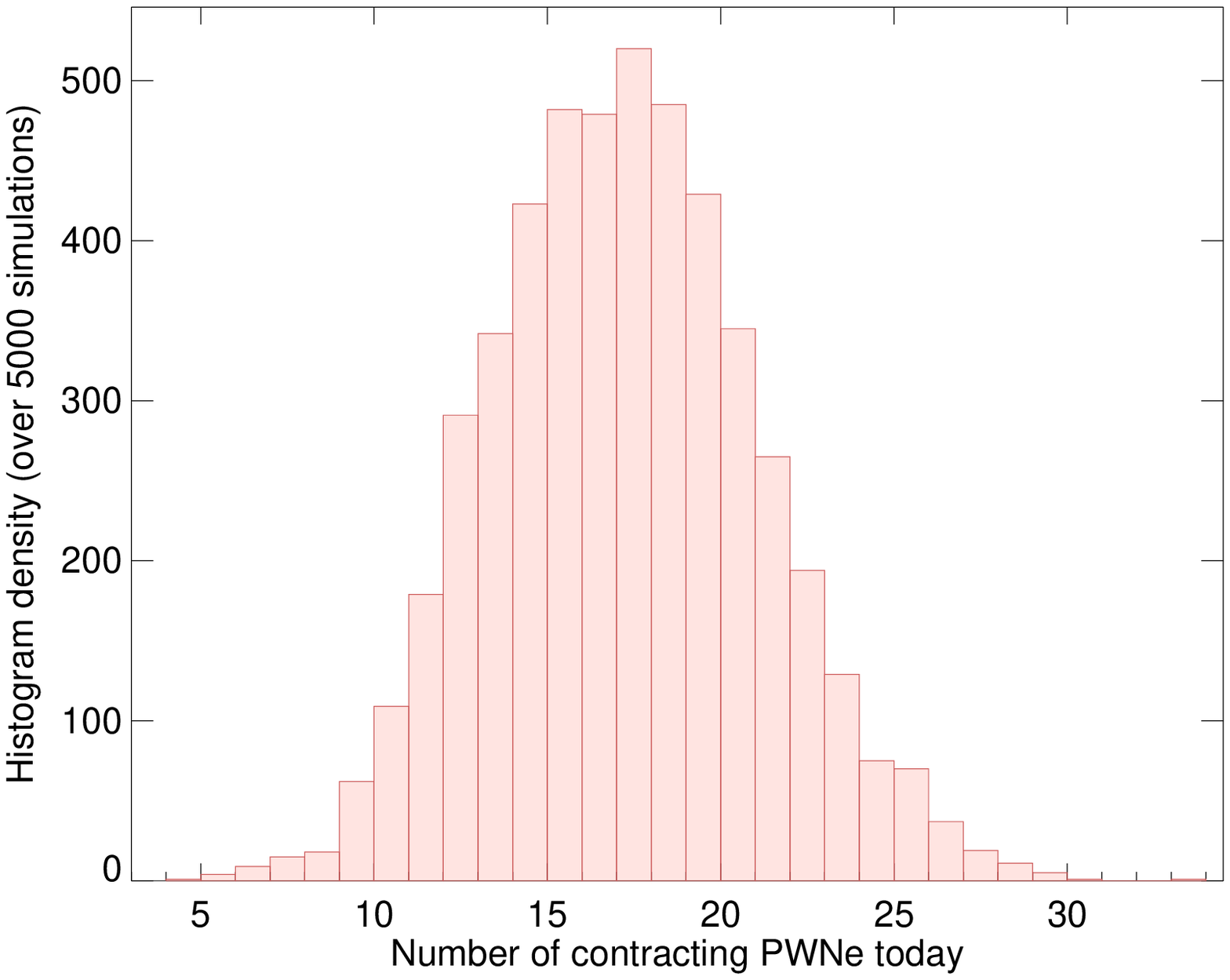}
    \caption{ Histogram density of the resulting number of Galactic PWNe that are superefficient and contracting today.}
    \label{dis2}
\end{figure}

\subsection{How many reverberating PWNe are there in the Galaxy?}

Using the same approach, we can calculate as well 
how many PWNe are currently in a reverberation stage. Particularly, how many of them are actually contracting.
Here, however, given that we have detailed models only for a range (although broad) of $L_0$, and that reverberation could also happen for pulsars having
initial spin down powers out of this range, our average should be considered a lower limit. 
Less luminous pulsars than those in Table 1 have in any case shorter reverberation periods as a whole, of a few hundred years at most, 
and missing them completely in our calculation is not expected to affect our average 
significantly. 
We find that out of the youngest 500 pulsars born in our Galaxy, 
those having $L_0$ values between $\sim 10^{35}$ and $\sim 10^{40}$ erg s$^{-1}$,
would lead to 16.7$\pm$3.8 currently contracting PWNe.
This is a relatively large number that 
promotes observational searches.

\section{Observational search of superefficient PWNe}

\subsection{May we have detected a superefficient PWN candidate already?}

Probably not. Since all X-ray sources in the Galaxy with luminosity $L_X > 1 \times 10^{35}$ erg s$^{-1}$ have been already
identified (e.g., \cite{vandenberg2012}),
we can safely conclude that if any PWN is currently in a superefficient stage, it must be a  relatively 
dim system.
However, out of the models studied, we do not find clear cases of PWNe for which
the spin-down power at the time of the maximum efficiency is $< 10^{34}$ erg s$^{-1}$
and the X-ray superefficiency is $<10$.
We thus conclude that observing PWNe not in superefficiency, but on the path towards superefficiency may be a better approach to test our theoretical predictions (see the next Section).
Before tackling this, we shall first confront our
prejudice and look for any possible
superefficient PWN candidate in  archival data.

Based on our theoretical computations, we hunted for superefficient PWNe candidates that may have already been detected adopting three different approaches. 
As a first check, promising candidates were selected by cross-matching young ($\tau_c\lesssim10$~kyr) pulsars with a relatively low spin-down power ($L_{sd}\lesssim10^{34}$ \lum) listed in the ATNF pulsar catalog (v.~1.59; see \cite{Manchester2005}) with unidentified point-like sources listed in the \xmm\ slew survey catalog \citep{Saxton2008}\footnote{The \xmm\ slew survey covers about 84\% of the whole sky with a sensitivity limit on the observed flux of $\sim1.2 \times 10^{-12}$ \flux\ in the 0.2--12~keV energy interval.} that might be associated with bright X-ray emission from the putative surrounding PWN. 
We did not find any pulsar fulfilling all the requisites above besides the known magnetars, high-magnetic-field pulsars and central compact objects. 

Alternatively, bright PWNe could exist around particularly faint radio pulsars which might have not been detected yet by means of radio pulsar surveys (and hence are not listed in the ATNF catalog). 
We thus searched for unidentified X-ray sources at low Galactic latitude ($|{\it b}|<4$\deg) listed in the \xmm\ slew survey catalog with a flux $F_X>10^{-11}$ \flux\ over the 0.2--12~keV energy band, and never observed using dedicated X-ray pointings. 
Notably, this criterion selects only three sources, XMMSL2\,J180205.6$-$251231 (at $b=-1.3$\deg), XMMSL2\,J123629.8$-$664549 (at $b= -3.9$\deg) and XMMSL2\,J194402.0$+$284451 (at $b=2.4$\deg). 
We requested follow-up X-ray and UV observations with the \emph{The Neil Gehrels Swift Observatory} \citep{Gehrels2004} for these sources, and the results are as follows.

\begin{table*}
\begin{center}
\caption{Multiwavelength properties of the two sources selected as possible superefficient PWNe candidates in the \xmm\ slew survey catalog, and detected in our recent \swift\ observations (see the text for details).} 
\label{tab:phot}
\resizebox{2.1\columnwidth}{!}{
\begin{tabular}{ccccccccc} \hline
Source name			& R.A., Decl. 				& $D$			& XRT count rate	& $N_{\rm H}$			& $\Gamma$	& $L_X$			& {\it Gaia} mag 	& UVW2 mag	\\ 
					& (J2000.0)				& (pc)			& (counts ks$^{-1}$)	& (10$^{21}$ cm$^{-2}$)	&			& (10$^{31}$ \lum)	& 				&	\\
\hline		
J123629.8$-$664549 	& 12$^h$36$^m$32$^s$.59, 	& 362(343)(384)	& $120\pm10$		& $3.7\pm1.5$			& $1.5\pm0.3$	& $14\pm2$		& 17.7			& $18.24\pm0.09$	 \\  \vspace{0.2cm}
					&  -66$^\circ$46'01''.6 		&				&				&					&			&				&				&	\\
J194402.0$+$284451	& 19$^h$44$^m$01$^s$.90,   	& 434(417)(453)	& $97\pm11$		& $1.0\pm0.9$			& $1.8\pm0.5$	& $9\pm1$		& 17.6			& $16.56\pm0.05$	 \\
					&  +28$^\circ$44'51''.6		&				&				&					&			&				&				&	\\				
\hline
\end{tabular}
}
\end{center}
\end{table*}

J180205.6$-$251231 was not detected by the X-ray Telescope (XRT) during our 1-ks observation on 2018 May 20 (observation ID: 010693001). We set a 3$\sigma$ upper limit of 0.01 counts~s$^{-1}$ on the background-subtracted count rate over the 0.3--10~keV energy range at the position quoted in the \xmm\ slew catalog. This rules it out as a PWN candidate in a superefficient phase. 
The \swift\ observation of J123629.8$-$664549 started on 2018 May 23 at 06:08:53 (UTC) and lasted 1.1~ks (obs. ID 0001069400). That of J194402.0$+$284451 started on 2018 May 31 at 12:29:42 (UTC) and lasted 0.8~ks (obs. ID 00010695002).
In both observations, the XRT was operated in the photon counting mode, and the Ultraviolet and Optical Telescope (UVOT) in image mode using the UVW2 filter.
The two sources were detected by both the XRT and the UVOT, and are also spatially coincident with sources in the {\it Gaia} release \citep{GAIA2018}. From these findings, we can infer their distances as well as the presence of a relatively bright UV/optical counterpart. Table~2 reports a summary of their properties.
Coordinates are taken from the second {\it Gaia} data release \citep{GAIA2018}. 
Distances were computed from the parallaxes measured with {\it Gaia} using a Bayesian method with exponentially decreasing volume density prior and accounting for the known systematic zero-point offset of -29~$\mu$as in the {\it Gaia} parallaxes. 
The 16th and 84th percentiles are reported in parentheses (see \citet{Bailer-Jones2018} for details). 
The XRT count rates and the X-ray luminosities refer to the 0.3--10~keV energy range.  Optical magnitudes are evaluated over the 3300--10500~\AA\ wavelength range. 
All quoted magnitudes are not corrected for absorption, and uncertainties on the UV values are quoted at a confidence level of 1$\sigma$.
We rule out that these sources could be good superefficient PWNe candidates owing to their low X-ray luminosity.

Finally, as a third search approach, we consider that our model predicts that the superefficient radiation mechanism may often yield to detectable gamma-ray emission (see Fig.~\ref{expectations}). 
However, on the one hand, cross-correlation between sources in the \xmm\ slew survey catalog and unidentified TeV sources listed in
the \emph{H.E.S.S.}~survey of the Galactic plane \citep{HESS-GPS-2018} did not yield any match. 
On the other hand, we found a few X-ray bright sources in the \xmm\ slew catalog (with $F_X>10^{-11}$ \flux\ in the 0.2--12~keV energy band) positionally coincident with unidentified gamma-ray sources listed in the preliminary \fermi\ Large Area Telescope (LAT) eight-year point source list.\footnote{\url{https://fermi.gsfc.nasa.gov/ssc/data/access/lat/fl8y/}}
However, the location of these sources are well above the Galactic plane (|{\it b}|>10\deg\ for all of them),
and are thus probably not related to any Galactic object. 
The deep all-sky X-ray survey that will be performed with the upcoming mission \emph{e-ROSITA} \citep{Merloni2012} is expected to lead to the discovery of several new fainter unidentified X-ray sources along the Galactic plane. Such a survey, coupled to the continuous monitoring of the Galactic plane in the GeV-TeV domain with
\fermi\ LAT and  \emph{H.E.S.S.}, will increase the chance to detect (fainter) Galactic superefficient PWNe candidates based on the above-mentioned criteria.

\section{Observational search of reverberating PWNe}

\subsection{Observing reverberating PWNe}

Given how unlikely it is that we have already detected a superefficient PWN, as well as because of the larger number of PWNe
that are expected to be in the reverberation phase now, it makes sense to search for observational signals of the latter. 
When identified, 
these PWNe will be 
running on the path towards their maximum efficiency, which will be expected to happen relatively soon (maybe just a few hundred years ahead).
Although the rate of change of observational features such as flux and spectrum may not be as large as it is in the superefficient phase,  
reverberating PWNe models provide predictions that might be at reach for current and/or forthcoming telescopes.

\begin{figure}
    \centering
    \includegraphics[width=0.49\textwidth]{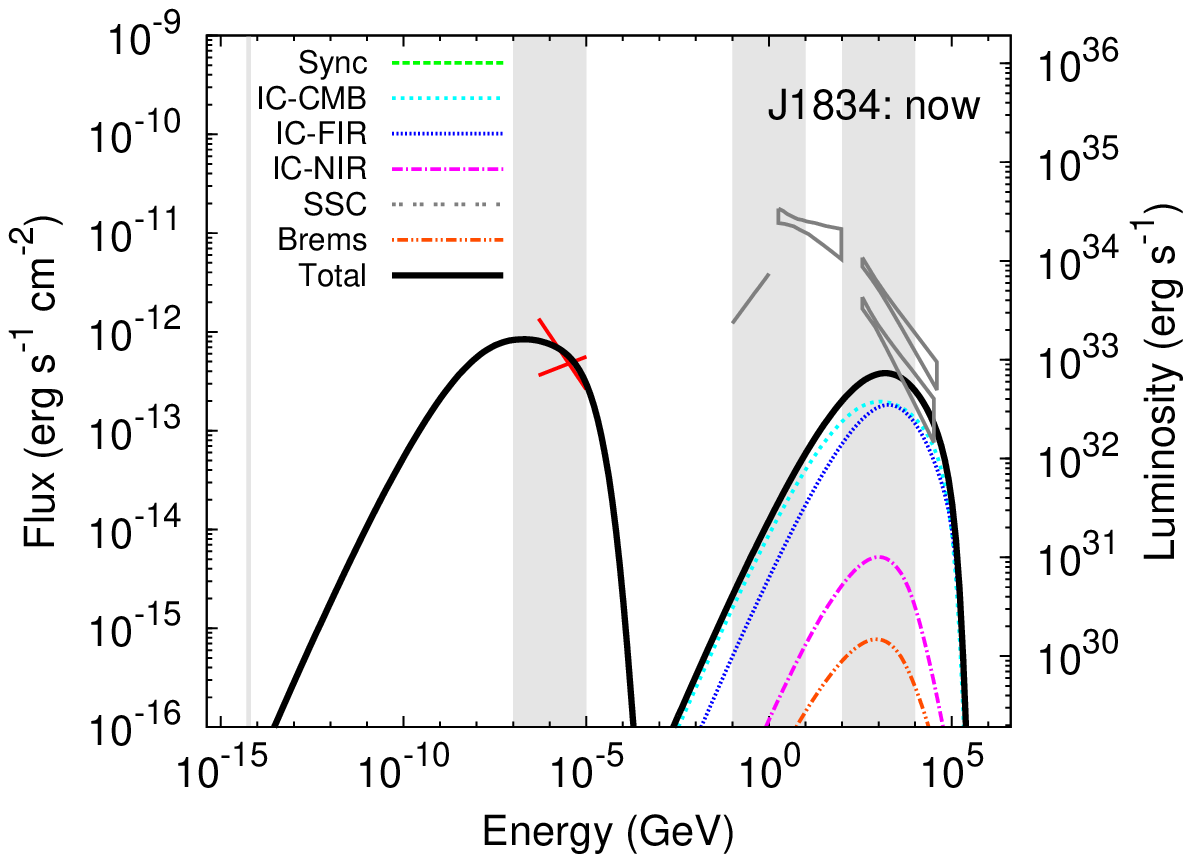}
    \includegraphics[width=0.49\textwidth]{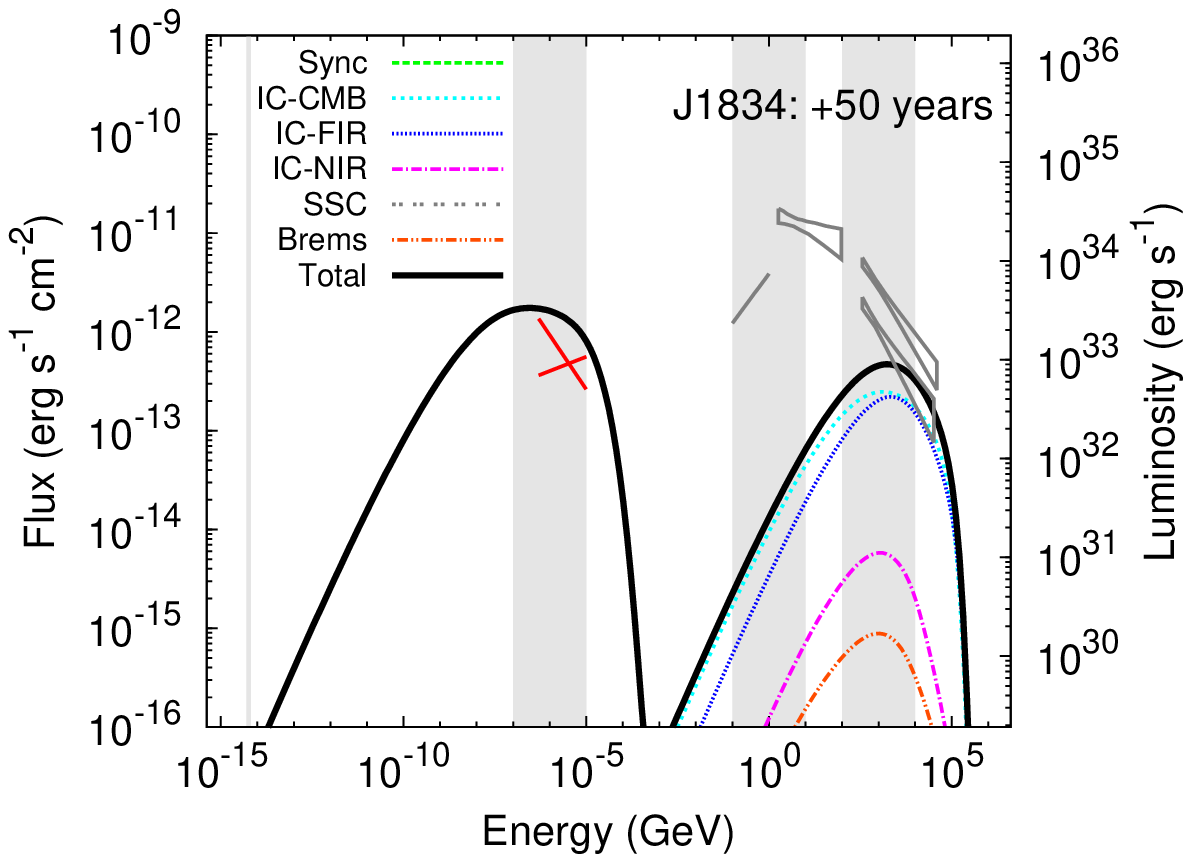}
    \caption{Spectra of the nebula J1834 now and in 50 years from now. The data is the same in both cases, and although not simultaneous they are obviously closer to
    the current time, so that in the second panel are put only for reference. 
    See text for discussion. The grey vertical lines and 3 grey rectangles from left to right represent the energy ranges considered for radio, X-ray, GeV and TeV, although data in the latter two ranges are only considered as upper limits --and the model parameters are thus taken to match them only to give a constraint on photon densities. }
    \label{SED-1834}
\end{figure}

Probably the best candidate for this study is the recently detected PWN around the magnetar J1834.9-0846 \citep{Younes2012,Younes2016}.
We have modelled this PWN in the same framework we are using in this work  before \citep{Torres2017}, as well as provided a detailed discussion regarding the 
dynamical situation, its age, and its birth supernova. 
We concluded there that this PWN can be rotationally-powered, despite having the largest efficiency known in X-rays, 
if it has recently entered into its reverberation phase. The deduced age today, 7970 years, is consistent with other dynamical 
features of the systems, its (non-)relation with W44, and the general energetics. 
We pose here that if J1834 is indeed reverberating, some changes in the following 50 years should be detectable. We start by describing them
in Fig. \ref{SED-1834}.
The upper panel of that figure 
represents the SED at the current time,  whereas the lower panel represents the time evolution of this SED as it will look 50 years into the future, so as to give a rough, preliminary idea of the changes expected.
Both the model and the origin of the (current) data set used for a comparison are detailed in \citet{Torres2017}, from where the model parameters were taken.
Briefly, the GeV data comes from \cite{Li2017} analysis of 6-years of {\it Fermi}-LAT data of observations of 
all magnetars, including J1834.
Having removed
the surrounding gamma-ray sources, and modelled out W41, the upper limits apply to the 
nebula directly. 
The TeV data represent the spatially coincident source
HESS J1834-087 \cite{Aharonian2006, Albert2006},
which may be unrelated to the PWN J1834, but originated in the interaction of particles accelerated in the W41 SNR with molecular clouds in the vicinity, e.g., 
\cite{Aharonian2006, Albert2006, Tian2007, Li2012, Castro2013,Abramowski2015}, similar to the case of W44 \citep{Uchiyama2012}, or W28 \citep{Hanabata2014}.
Thus, we shall consider the TeV data from HESS J1834-87 to represent upper limits 
to the putative TeV emission of the magnetar nebula. 
The photon energy density in the infrared and far-infrared are assumed such that the model would produce the maximum TeV emission still in agreement with these observations.
For further details on the data used and the model itself see \cite{Torres2017}.

\begin{figure}
    \centering
    \includegraphics[width=0.49\textwidth]{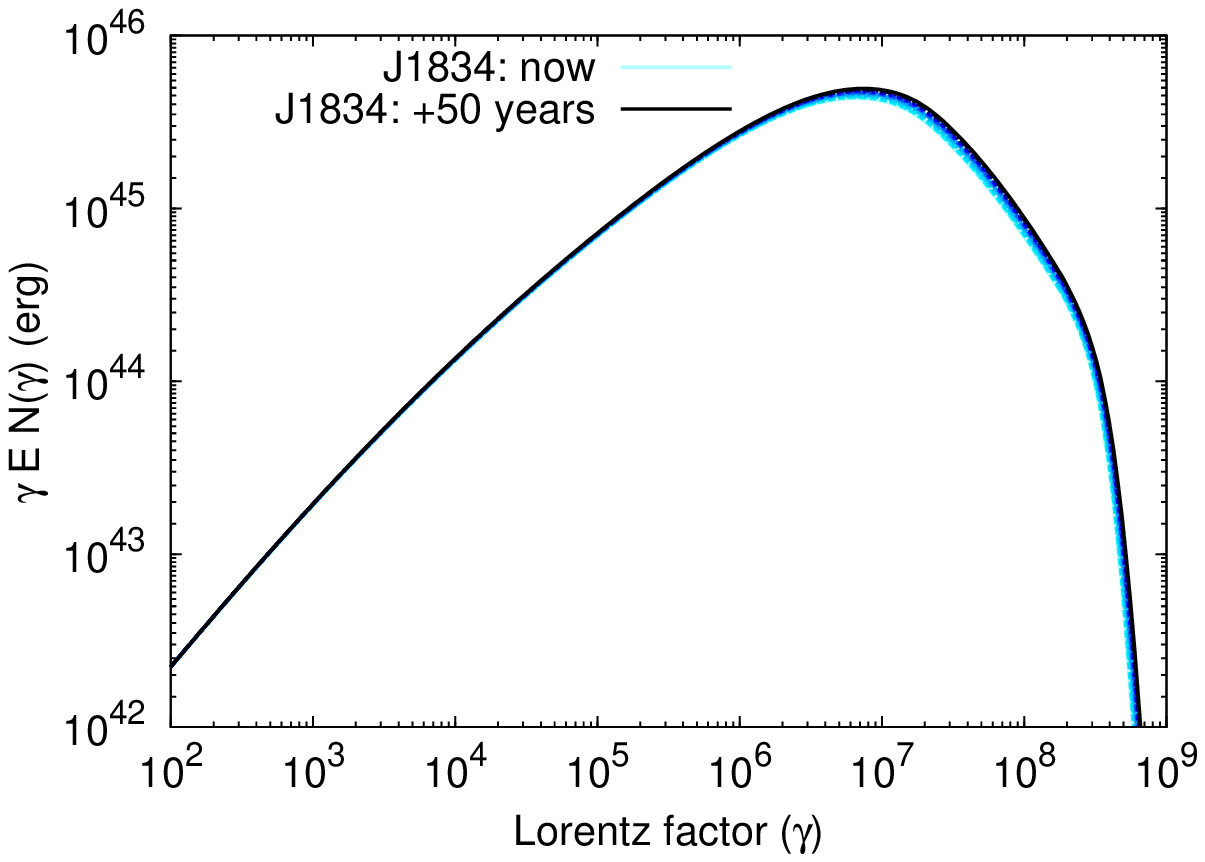}
    \includegraphics[width=0.49\textwidth]{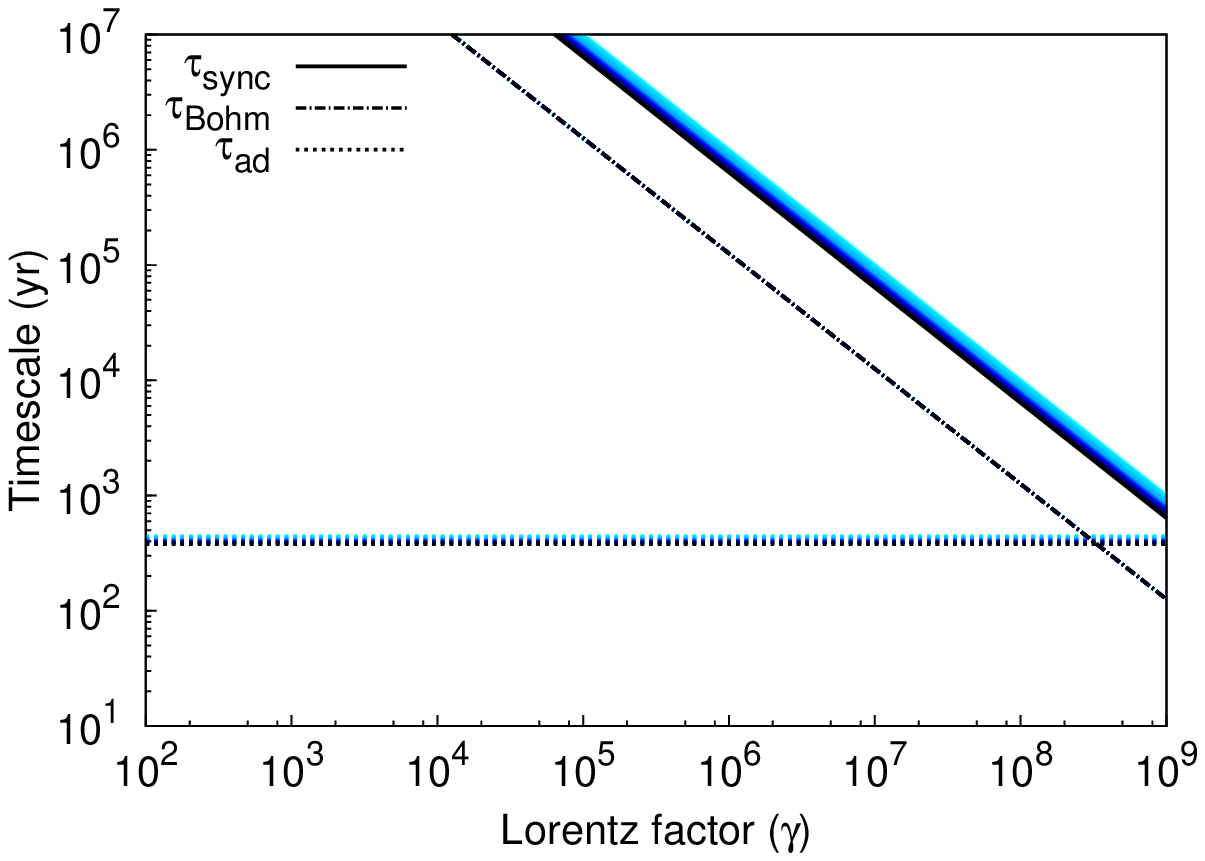}
    \caption{Evolution of the electron spectrum as well as the timescales of electron losses via synchrotron radiation, Bohm diffusion, and the adiabatic gain of energy due to the contraction of the system. The color scale (from light to dark, the same used in the next figures) shows the time evolution, in steps of 5 years, although curves are close in this scale to individually distinguish them. 
}
    \label{electrons-1834}
\end{figure}

The flux given in the plot of Fig. \ref{SED-1834}
is obtained just as distance-diluted luminosity predictions, so that possible effects along propagation (e.g., absorption in the optical or UV) are not considered. 
It is clear that J1834.9-0846 will become visibly brighter 50 years from now than what currently is.
The heating of electrons produced by the compression of the PWN is the reason behind this behavior. This is shown in Fig. \ref{electrons-1834}
where the color scale shows the time evolution of both electrons and losses, in steps of 5 years. 
Differences are almost imperceptible in this scale, except for the region of the maxima, where the burning is more extreme as time goes by.

\begin{figure*}
    \centering
    \includegraphics[width=0.49\textwidth]{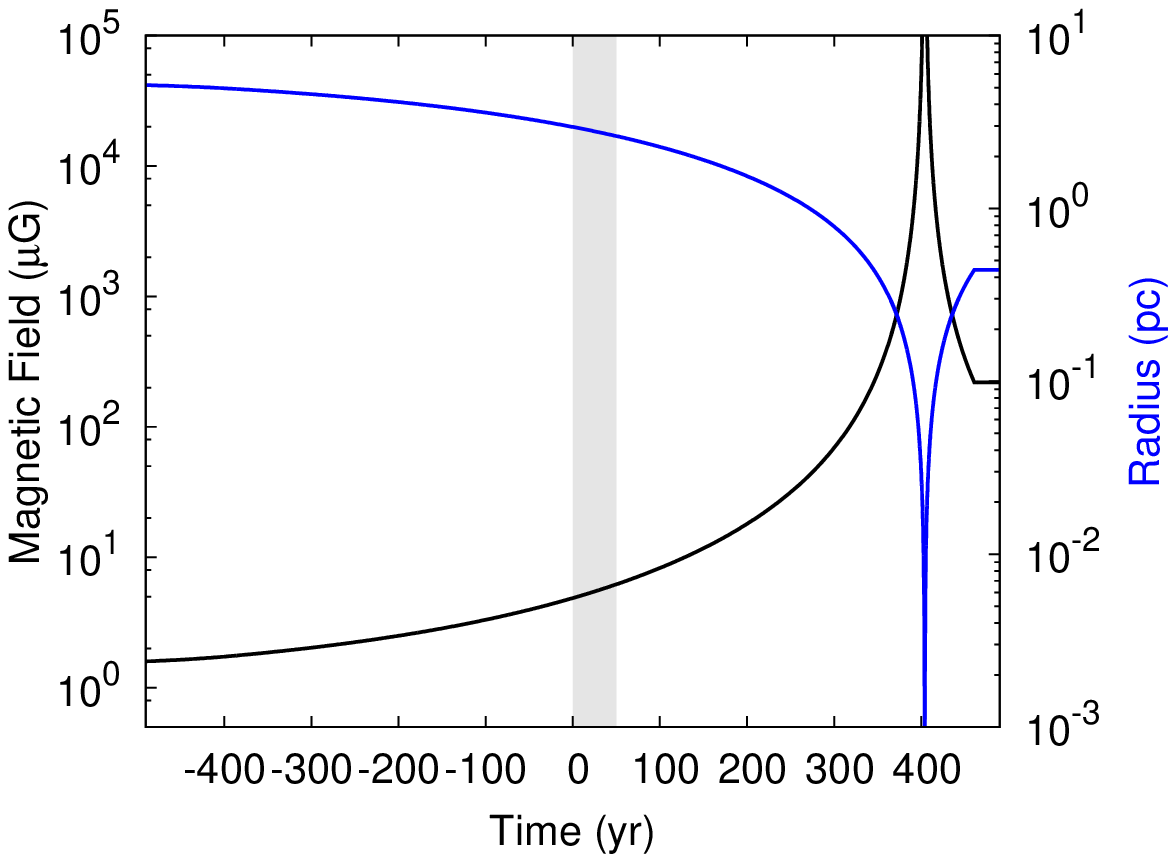}
      \includegraphics[width=0.49\textwidth]{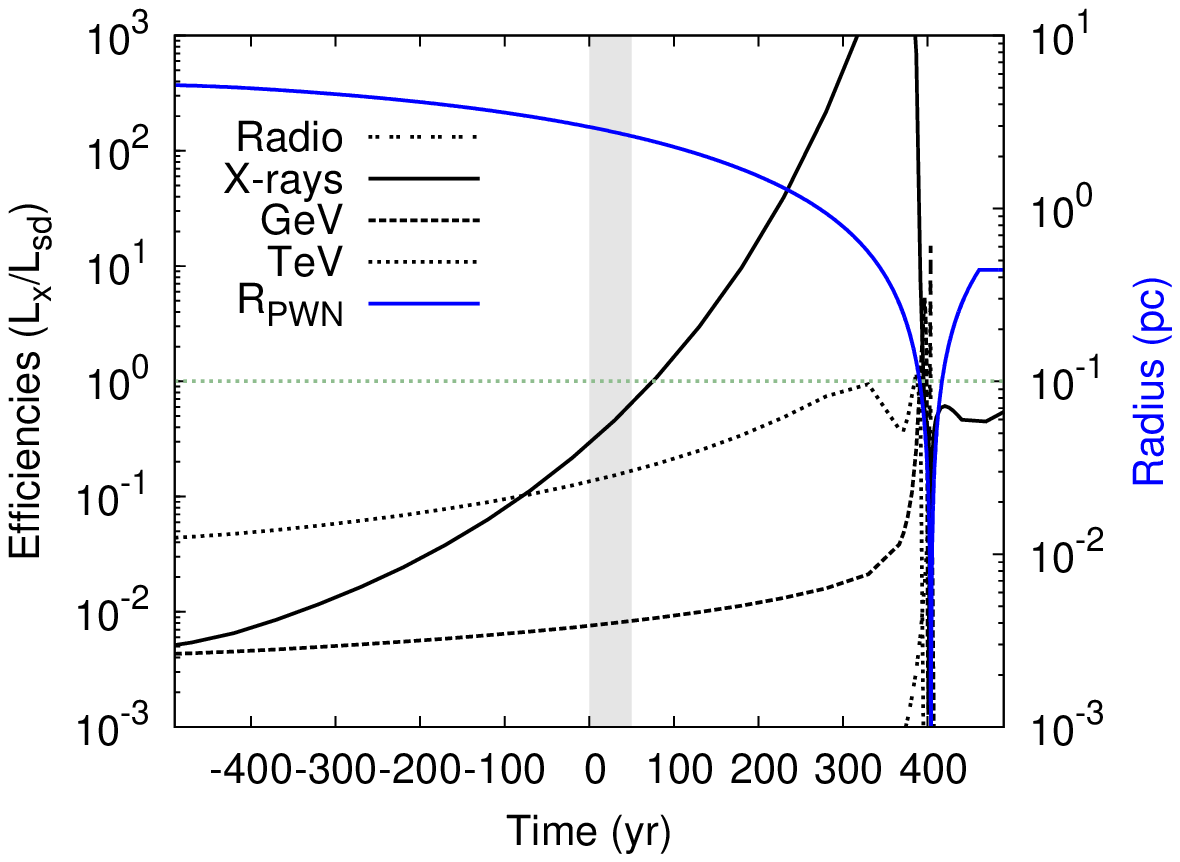}\\
             \includegraphics[width=0.49\textwidth]{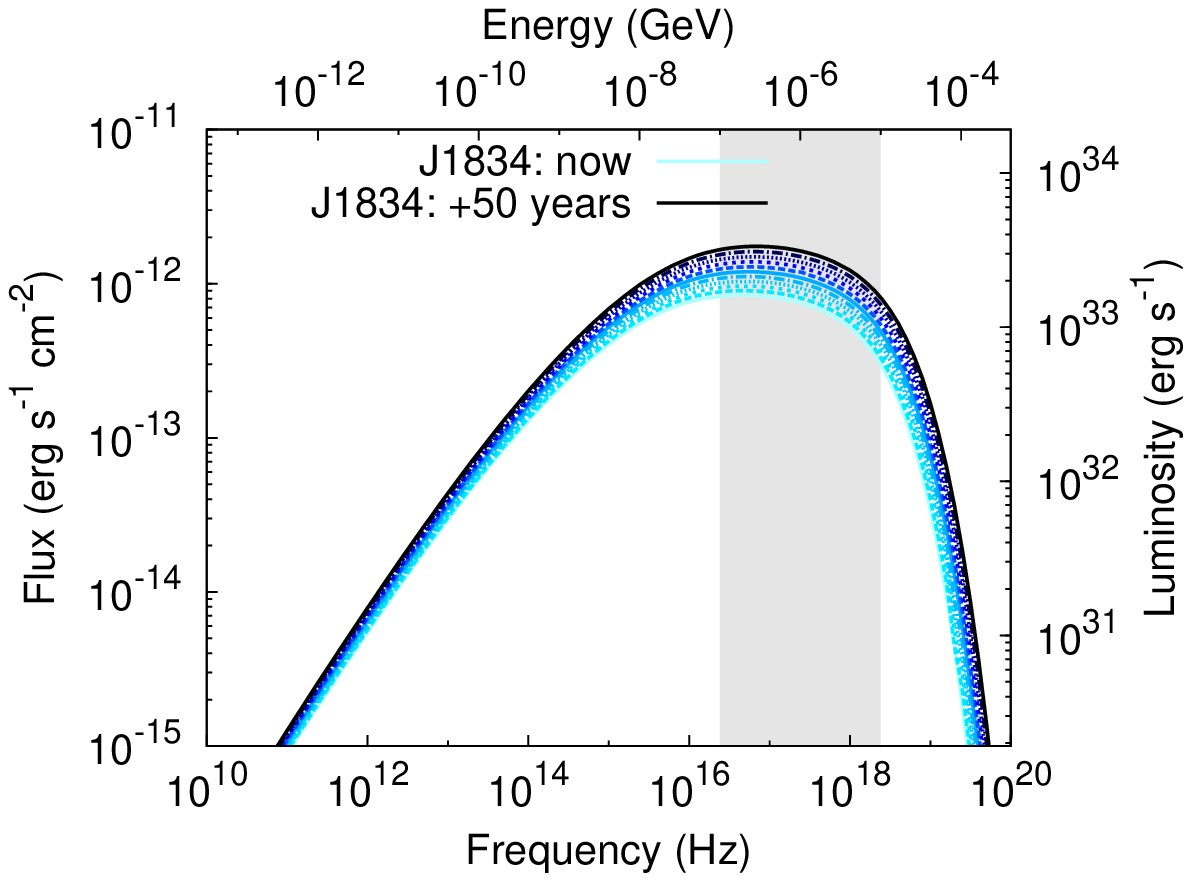}
       \includegraphics[width=0.49\textwidth]{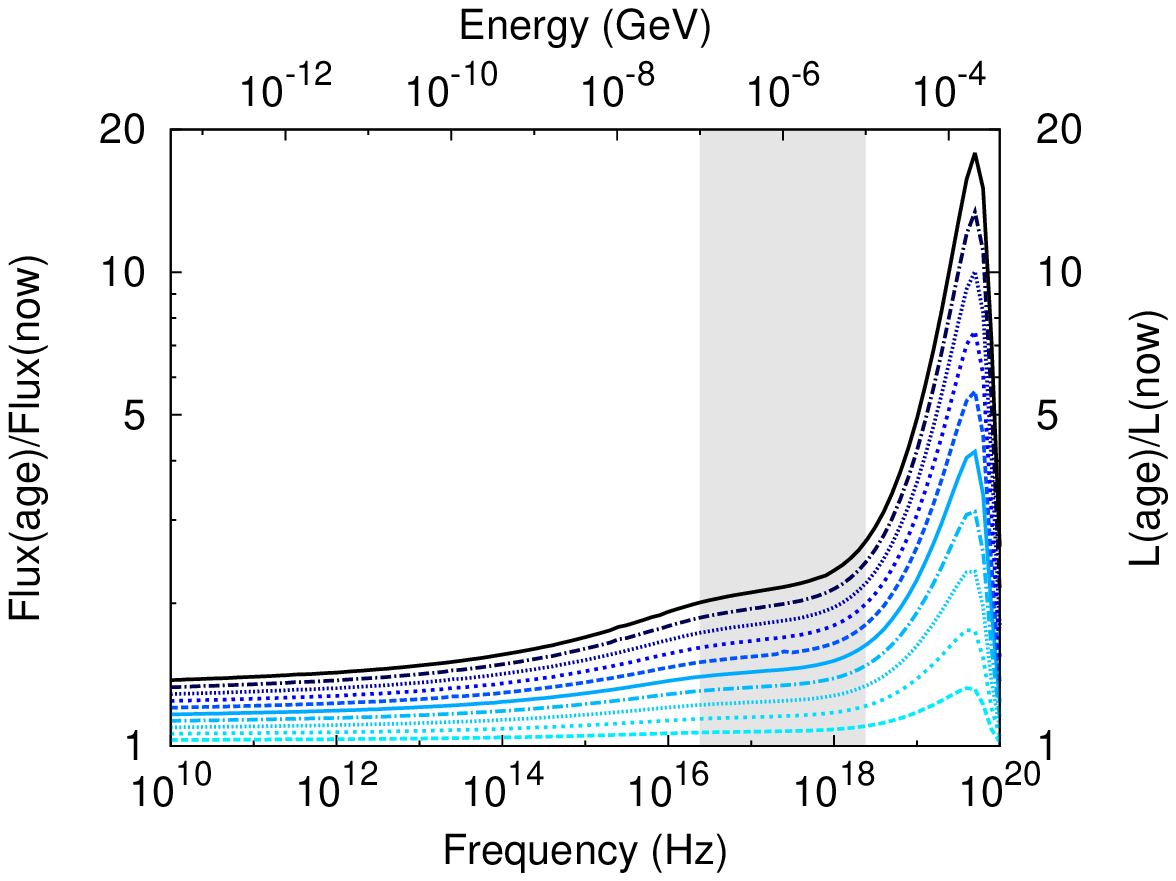}
    \caption{Top: Contraction and enhancement of the magnetic field due to reverberation process.  The label 0 in the x-axis represent the current time, with positive (negative) values 
    following the evolution towards the future (past).  The top-right panel shows the evolution of the efficiencies of the emission along time. The grey shadows in the top panels
    represent the sub-sequent 50 years of evolution.
    The bottom plots emphasizes the changes expected in the X-ray spectra. The grey shadows in the bottom panels show the region of 0.1--10 keV.
     }
    \label{b-R}
\end{figure*}

{  The two upper panels of Fig. \ref{b-R} } show the subsequent evolution of the PWN in context. The first panel shows the evolution of the radius 
and the enhancement of the magnetic field due to the reverberation process. 
The top-left panel shows an extended period of time covering the whole process of contraction and re-expansion. The grey-shaded region zooms in a period of 50 years starting in 2017. The magnetic field and radius of the nebula increases and decreases, respectively, as a shallow power law in this period, both with variations well  within a factor of 2. 
The top-right panel shows the evolution of the efficiency. 
   It is clear that despite the minimal changes in radius and magnetic field described above, the X-ray efficiency (and luminosity) is the most affected, also reaching a change up to a factor of 2 within the next 50 years. 
  This is better seen in the evolution of the synchrotron spectrum and in the evolution of the ratio of luminosities (bottom panels) normalized to the one attained today, and with the same color coding 
  in steps of 5 years.
  From the last panel we also see that the harder the X-ray energies of the measurement, the better, since the synchrotron emission is more affected at the largest electron energies.

\subsection{Are the changes predicted for the J1834 PWN observable?}

Our model makes specific predictions on the time evolution of the X-ray emission properties of the nebula around J1834, which can be tested against observations. As an illustrative case, we verified if the expected flux increase of the nebula could be detectable with the \xmm\ EPIC pn instrument \citep{Struder2001} over a timescale of 10\,yrs between the epoch of the last observations in 2014 \citep{Younes2016} and 2024. We assumed an uncertainty of 10\% on each data point in the theoretically predicted SEDs of the nebula at the two epochs, and fitted a power law model to the data points over the 0.3--10\,keV energy range. This allowed us to estimate the expected photon index and the intrinsic (i.e. unabsorbed) flux of the nebula at the two epochs. We note that assuming a slightly smaller and/or larger uncertainty on the SED data points yielded values for the photon index and the flux fully consistent with each other. On the one hand, the SEDs show no substantial change in the spectral slope, in agreement with what was found by \cite{Younes2016} when comparing with earlier observations. On the other hand, they imply a flux increase of $\Delta F_X\simeq4\times10^{-13}$~\flux\ between 2014 and 2024.

We carried out simulations using the \textsc{fakeit} tool within \textsc{xspec}. 
We assumed for simplicity no dependence of the spectral shape of the nebula on the location, i.e. we considered the case of a homogenous emitting nebula. 
We accounted for the absorption by neutral interstellar material along the line of sight adopting the value for the hydrogen column density reported by \cite{Younes2016} assuming the photoionization cross-sections from \cite{Verner1996} and the chemical abundances from \cite{Wilms2000}, i.e. $N_{\rm H}=8\times10^{22}$\,cm$^{-2}$.
We used the latest versions of the response files for the pn in the extended full-frame configuration and with medium optical blocking filter in place
(i.e. the same setup adopted in 2014). Background spectra were extracted from the pn data acquired in 2014 using the same prescriptions outlined by \cite{Younes2016}. 
We simulated a 130~ks-spectrum for the 2014 epoch. This corresponds to the effective net exposure time of the \xmm\ EPIC pn observations at the same epoch (see table\,1 by \cite{Younes2016}). Different exposure times were adopted instead to simulate the spectrum as would be acquired in 2024. We then modelled each spectrum with an absorbed power law, holding $N_{\rm H}$ fixed to the above-mentioned value, and allowing the photon index and the normalization to vary. 

Spectral fitting as a function of time shows that an 80-ks exposure with the EPIC pn in 2024 would allow us to measure the X-ray flux of the nebula with enough accuracy to detect the predicted flux increase on a 10\,yr-timescale at a confidence level $\gtrsim4\sigma$.
This measurement would be more precise (or done in a shorter time) if observed with more sensitive instruments than EPIC pn. Such
a relatively short observation would confirm the PWN is indeed reverberating, and open the gate to testing particular details of the model.

In particular, to test whether the predicted rate of flux increase could be detected on timescales as short as three years with future X-ray imaging instruments, we simulated the spectra of the nebula as would be acquired by the Wide Field Imager (henceforth WFI; \cite{Meidinger2017}) on board the \textit{Advanced Telescope for High Energy Astrophysics} (\textit{Athena}), as well as by the Spectroscopic Focusing Array (SFA; \cite{Zand2019}) on board the \textit{enhanced X-ray Timing and Polarimetry mission} (\textit{eXTP}), following the procedure outlined above.
The fluxes expected at four different epochs over the planned mission durations (2023--2028 for \textit{eXTP} and 2028--2033 for \textit{Athena}) are listed in the second column of Table~\ref{tab:simul}. Our simulations are aimed at detecting the rate in the flux increase predicted by our models, $\simeq(4-5)\times10^{-14}$~\flux~yr$^{-1}$ on a timescale of 3 years between 2024 and 2033.

\begin{table}
\small
\begin{center}
\caption{Predicted time evolution of the nebula unabsorbed flux over the 0.3--10\,keV energy band, and results of our simulations with \textit{eXTP} and \textit{Athena}. 1$\sigma$ uncertainties are reported on the fluxes
measurable at the different epochs with 100-ks exposures (see the text for details).} 
\label{tab:simul}
\resizebox{1.\columnwidth}{!}{
\begin{tabular}{cccc} \hline
Year 	    	& Predicted flux 	& Instrument		& Measurable flux 			\\
         	& (10$^{-12}$ \flux)   &				& (10$^{-12}$ \flux)			\\
\hline 
\vspace{0.15cm}			
2024		& 2.69			& \textit{eXTP} SFA	& $2.67\pm0.04$ 			\\ \vspace{0.15cm}
2027		& 2.82			&				& $2.79\pm0.04$ 			\\ \vspace{0.15cm}
2030		& 2.95			& \textit{Athena} WFI	& $2.92\pm0.03$ 			\\ \vspace{0.15cm}
2033		& 3.10			&				& $3.09\pm0.03$ 			\\ 
\hline
\end{tabular}
}
\end{center}
\end{table}

\begin{figure}
    \centering
    \includegraphics[width=0.47\textwidth]{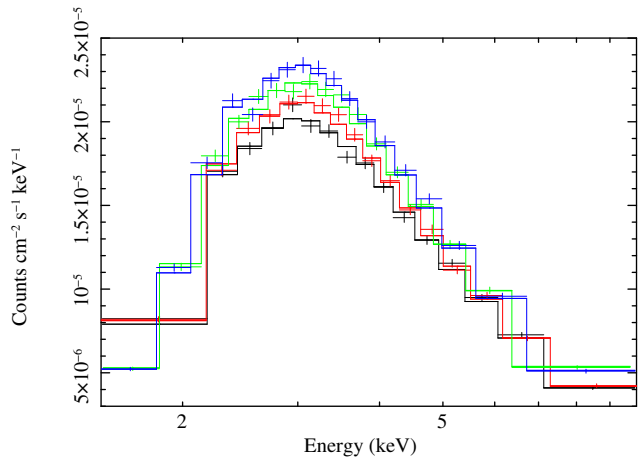}
    \caption{Simulated spectra of J1834's nebula at four different epochs (see the text for details). Black: in 2024; red: in 2027; green: in 2030; blue: in 2033. For plotting purpose, spectra have been rebinned and are shown  over the 1.5--10 keV energy range.}
    \label{athena}
\end{figure}

We employed the latest versions (2017 November) of the response files for the nominal WFI mirror configuration with 19 rows and without external light blocking filter. 
We also accounted for the expected instrumental and diffuse background level using the publicly available, latest version (2018 July) of the background files for extended sources (\url{http://www.mpe.mpg.de/ATHENA-WFI/response_matrices.html}).
Similarly, we adopted the latest versions of the response, auxiliary and background files to simulate the spectra related to the SFA (\url{https://www.isdc.unige.ch/extp/response-files.html}).
%
%
Spectral fitting as a function of time shows that 100-ks effective exposures would allow us to measure the X-ray flux of the nebula with enough accuracy to detect the predicted flux increase on a 3\,yr-timescale at a confidence level $\gtrsim3\sigma$ (see Table~\ref{tab:simul}). 
The simulated spectra at the different epochs are shown in Figure~\ref{athena}.

\section{Concluding remarks}

In this work we have studied, from both a theoretical and an observational perspective, reverberating and superefficient PWNe.
Based on a series of systematically studied PWNe models we quantified the 
duration and significance of these periods in the PWN evolution.
Our theoretical results for 
the periods in which the luminosity in a given band (for instance, in X-rays) exceeds the spin-down power
are mainly summarized in Fig. \ref{expectations}.
There, for a series of models based 
on varying the initial spin down power of specific pulsars, we computed the 
maximum efficiency in different bands and the corresponding duration of the superefficiency interval as a function of the pulsar energetics 
and the minimum radius attained. 
The efficiencies are larger for the most energetic pulsars, but the duration of these superefficient periods are smaller. 
Pulsars being more energetic, they are able to sustain the pressure of the medium earlier and their PWNe correspondingly 
bounce before.
The most efficient pulsars are also those attaining the smaller minimum PWNe radii.

{  The possible existence of superefficient states is in quite general terms independent of model details, since it is based on 
a conceptual energetic argument in which the environment adds energy to the system due to compression, see \cite{Torres2018b}. However, how large it is
(equivalently, which is the amount of maximal compression) will depend on the reality of the assumptions made for the dynamical
evolution.}

{  
The closest model to ours is that of \cite{Gelfand2009}. 
In particular, regarding the dynamical evolution assumptions both models are comparable.
And we notice that they also found large compressions:
See, for instance, Figure 2 of \cite{Gelfand2009}, in which they show
a compression of a PWN by a factor of 16, even when they are considering as baseline the parameters used in Model A in the paper by \cite{Blondin2001}, where the $L_0$ is at the largest level,  10$^{40}$ erg s$^{-1}$ (with other parameters, e.g., ISM medium density, $\omega, E_{sn}, M_{ej}$, etc. being similar to those assumed here). 
Or see also how these large compressions appear in the case for the PWN CTA 1 as well; see figure 6 of \cite{Aliu2013}, where a factor of maximal compression of at least 100 is also shown.
}

{  
However, we caveat that large compression factors beyond 100 may seem excessive, particularly when compared with hydrodynamical (HD) simulations (see e.g., \cite{Bucciantini2004})
and may probably come from model simplifications. We recall that our models are 1-dimensional (thus, spherical, homogenous), which is not generally the case in reality.
For instance, Rayleigh-Taylor mixing, see \cite{Bucciantini2004b,Gelfand2009}, activated when the $P_{pwn}>P_{snr}(R_{pwn})$ during free expansion, may later damp coherent compression to this extent. HD and magneto HD simulations of the reverberation period 
may help understand this dichotomy. However, HD simulations may also be caveated by the fact that the radiative losses of particles are usually not consistently considered (or, simply, not considered at all) in the computation of the PWN pressure. This may lead to an overestimate of the PWN pressure, and thus to diminish the compression factors, since particularly in the reverberation phase, radiative losses of particles can be large. 
Further analysis are needed to clarify these points.}

{  
Interestingly, though, the predictions studied for J1834, on which we also comment below, should not be significantly affected by a reduction of a minimum radius acquired in the reverberation process. This PWN is supposed to be at the start of the reverberation phase, yet far from the maximal compression achievable -a situation that will also be valid in 50 years from now.
Also, given the very small duration of the highest-compression regime along the evolution, the duration of the total reverberation process should not change much, if at all. 
These  provide confidence that the simulations results for the number 
of contracting PWNe should also be in a reasonable range.}

Using our models and a Monte Carlo simulation of the youngest 500 pulsars born in our Galaxy, 
in accordance to population synthesis models, 
we were able to
determine that 
1.8 $\pm$ 1.3 PWNe should be superefficient today, and at least 
16.7 $\pm$ 3.8 should be currently contracting.
Of course these estimations (as far as we are aware, the first of their class) 
delineate the ballpark of what is reasonable to expect as a number of contracting systems in their different phases.
Caveats coming from a (still relatively) limited sample of detailed models on which the estimates are based (40 models were studied here), and 
the model limitations themselves should be bore in mind.
In particular, it is useful to note that the assumption of fixing the mass of the ejecta 
in these PWNe to that of the anchor case used (ultimately done in the lack of better knowledge) may not necessarily turn out to be 
the best approach in some cases.
For instance, models with relatively low $L_0$ and high ejecta mass (based for instance on G54) might not realize. 
Being aware of such uncertainty, however, we preferred to have a systematic approach 
for choosing the PWNe models used.

A direct observational search was also conducted.
Analysis of X-ray detections known till now from the
{\it XMM-Newton} Slew Survey did not identify clear candidates of superefficient PWNe.
In fact, three interesting 
sources were signalled in our investigation, but additional observations done 
with {\it Swift-XRT}
rule them out as possible superefficient PWNe.
We concluded that we have probably not detected a superefficient PWNe yet, and that probably none currently exist
in the Galaxy (unless of course we are unlucky enough so that a superefficient PWNe lies in the 16\% of the sky not
covered by the {\it XMM-Newton} Slew Survey).

However, linked to the reverberation phenomena, superefficient PWNe will inexorably exist 
in the future.
While we wait, reverberation itself can be detected. 
We have shown this using the specific case of J1834. 
In less than 10 years from now, we have shown that EPIC pn should be able to tell the difference in flux between the flux measured 
for this source in 2016 and before and the one obtained then. The latter should be larger, corresponding to the enhancement produced by the contraction.
If this is confirmed, we have also shown that such source will become an ideal laboratory for instruments such as 
{\it eXTP} and {\it Athena}, since they will be able to measure the rate of X-ray increase towards superefficiency already in 3-years steps, 
allowing for 
a more detailed theoretical comparison.

\section*{Acknowledgments}

We acknowledge the support from the Chinese Scholarship Council, The National Key Research and Development Program of China (2016YFA0400800), the Spanish grants AYA2015-71042-P, SGR2017-1383, AYA2017-92402-EXP, iLink 2017-1238, and the National Natural Science Foundation of China via NSFC-11473027, NSFC-11503078, NSFC-11673013, NSFC-11733009, XTP project XDA 04060604 and the Strategic Priority Research Program ``The Emergence of Cosmological Structures" of the Chinese Academy of Sciences, Grant No. XDB09000000. 
We acknowledge J. Martin, J. Pons, B. Olmi, and M. Mezcua for discussions, and referee R. Bandiera for insightful remarks.


\label{lastpage}
\end{document}